\documentclass{article}
\usepackage{amsfonts,amsmath,amsxtra,graphicx}
\usepackage{wrapfig}
\usepackage{mathtext}
\usepackage[T2A]{fontenc}
\usepackage{amsmath,amsfonts,amsxtra, mathtools}
\usepackage{amssymb}
\usepackage{amscd}
\usepackage[usenames,dvipsnames]{color}
\usepackage{color}
\usepackage{euscript}
\usepackage[matrix,arrow,curve]{xy}
\usepackage{mathrsfs}
\usepackage{dsfont}
\usepackage{array}
\usepackage{indentfirst}
\usepackage{textcomp}

\def\be{\begin{eqnarray}}
\def\ee{\end{eqnarray}}
\def\nn{\nonumber}

\def\0{\emptyset}
\def\p{\partial}

\def\Tr{{\rm Tr}\,}
\def\l[{\phantom.[}

\def\X{w}

\def\lm{\limits}

\def \A {\mathcal{A}}

\def \IC {\mathbb{C}}
\def \IZ {\mathbb{Z}}
\def \IR {\mathbb{R}}

\def \wt {\mathop{\mathfrak{wr}}\nolimits}
\def \Pexp {\mathop{{\cal P}{\rm exp}\;}}
\def \wt {\mathop{\mathfrak{wr}}}

\hoffset= - 1.0in         
\numberwithin{equation}{section}

\begin{document}
\title{{\bf {Wall Crossing Invariants:\\
from quantum mechanics to knots} \vspace{.2cm}}
\author{{\bf D.Galakhov$^{a,b,}$}\footnote{galakhov@itep.ru; galakhov@physics.rutgers.edu}, {\bf A. Mironov$^{c,a,d,}$}\footnote{mironov@itep.ru; mironov@lpi.ru} and {\bf A. Morozov$^{a,d,}$}\thanks{morozov@itep.ru}}
\date{ }
}

\maketitle

\vspace{-6.0cm}

\begin{center}
\hfill FIAN/TD-14/14\\
\hfill ITEP/TH-28/14\\
\end{center}

\vspace{4.2cm}

\begin{center}
$^a$ {\small {\it ITEP, Moscow 117218, Russia}}\\
$^b$ {\small {\it NHETC and Department of Physics and Astronomy, Rutgers University,
Piscataway, NJ 08855-0849, USA }}\\
$^c$ {\small {\it Lebedev Physics Institute, Moscow 119991, Russia}}\\
$^d$ {\small {\it National Research Nuclear University MEPhI, Moscow 115409, Russia }}\\
\end{center}

\vspace{1cm}

\begin{abstract}
We offer a pedestrian level review of the wall-crossing invariants.
The story begins from the scattering theory in quantum mechanics
where the spectrum reshuffling can be related to permutations
of $S$-matrices.
In non-trivial situations, starting from spin chains and matrix models,
the  $S$-matrices are operator-valued and their algebra is
described in terms of ${\cal R}$- and mixing (Racah) ${\cal U}$-matrices.
Then, the Kontsevich-Soibelman (KS) invariants are nothing but the standard
knot invariants made out of these data within the Reshetikhin-Turaev-Witten approach.
The ${\cal R}$ and Racah matrices acquire a relatively universal form
in the quasiclassical limit, where the basic reshufflings with the
change of moduli are those of the Stokes line.
Natural from this point of view are matrices provided by the modular
transformations of conformal blocks
(with the usual identification ${\cal R}=T$ and ${\cal U}=S$), and in the simplest case
of the first degenerate field $(2,1)$, when the conformal blocks satisfy
a second order Shr\"odinger-like equation, the invariants
coincide with the Jones ($N=2$) invariants of the associated knots.
Another possibility to construct knot invariants is to realize the cluster coordinates associated with
reshufflings of the Stokes lines immediately in terms of check-operators acting on the solutions to the Knizhnik-Zamolodchikov equations. Then, the ${\cal R}$-matrices are realized as products of successive mutations in the cluster algebra and are manifestly described in terms of quantum dilogarithms ultimately leading to the Hikami construction of knot invariants.
\end{abstract}

\tableofcontents

\section{Introduction}

The string theory approach to any problem is to consider it together
with all possible deformations and as a particular representation of some
general structure appearing in many other, seemingly unrelated problems
in other fields of science.
One of the fresh application of this approach is the study of the wall-crossing
phenomena (phase transitions) and associated invariants, which remain the same
after the reshuffling.
The outcome of this study is that the Kontsevich-Soibelman (KS) invariants \cite{KS} found so far
on this way, are probably not that new:
they belong to an old class of invariants of the Reshetikhin-Turaev-Witten type, of which the most well-known are {\it knot invariants} \cite{Wit,RT}.
At the same time, what naturally arises in wall-crossing problems,
are quantum ${\cal R}$-matrices in representations less trivial than the Verma modules
of $SU_q(N)$, and this can further stimulate the study of knot invariants
in non-trivial representations.

An archetypical example of the wall-crossing is the spectrum dependence on the scattering potential in quantum mechanics.
Consider a particle in the infinite well with some localized potential,
for example:
\be
\Big(-\p_x^2 + u\delta(x) - k^2\Big)\psi(x) = 0, \ \ \ \ \
\psi(-L_1) = \psi(L_2) = 0
\ee
The spectrum $k(u)$ is defined by the spectral equation
\be
\sin\Big(kL_1+kL_2\Big) - \frac{u}{k}\sin(kL_1)\sin(kL_2) = 0
\ee
and changes from the set
\be
k_n=\frac{\pi n}{L_1+L_2} \ \ \ \ \ {\rm at} \ \ \ \ u=0
\label{speu0}
\ee
to a union of two sets\footnote{For $u>{L_1+L_2\over L_1L_2}$ there are also two bound states with $k = \pm i \kappa$, where $\kappa$ solves the equation
\be
\sinh\Big(\kappa L_1+\kappa L_2\Big) - \frac{u}{\kappa}\sinh(\kappa L_1)\sinh(\kappa L_2)=0\nonumber
\ee
}
\be
k_{n}^I = \frac{\pi n}{L_1}, \ \ \  k_n^{II}= \frac{\pi n}{L_2}\ \ \ \ \ {\rm at} \ \ \ \ u=\infty
\label{speuinf}
\ee
The smooth evolution with $u$ is shown in Fig.1,
but the net result is the rather radical reshuffling of (\ref{speu0}) into (\ref{speuinf}).
The task can be to study this reshuffling and to ask if there are quantities
that remain the same after the reshuffling.

\begin{figure}[htbp]
\begin{center}
\includegraphics[scale=0.1]{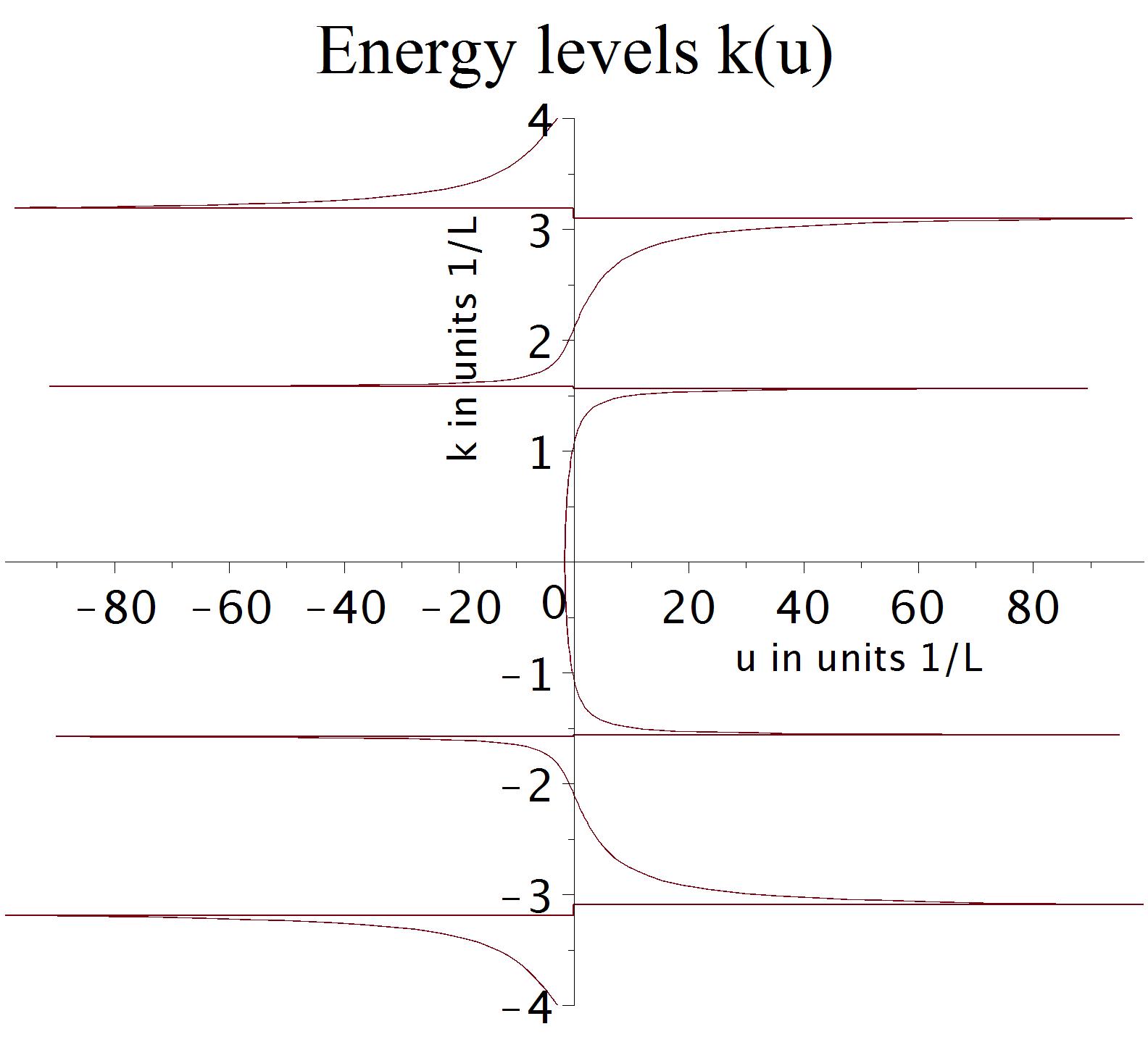}
\caption{The picture of energy levels as function of $u$ at $L_1=2L_2:=2L$.\label{energy}}
\end{center}
\end{figure}

The question is actually uninvestigated, but its more sophisticated versions
were studied and {\it some} invariants were revealed (it is, however, unclear if
they reduce to triviality in this original problem).

The point is that there is nothing special about the $\delta$-function potential:
the pattern remains the same for arbitrary barrier, vanishing in the vicinity of
the box walls.
Any such a problem is described in terms of the $2\times 2$ scattering matrix
\be
{\cal S}:   \ \ \ \left.e^{\pm ikx}\right|_{x \ {\rm near}\ -L_1}
\longrightarrow \left.\alpha_\pm(k) e^{\pm ikx} + \beta_\pm(k)e^{\mp ikx}\right|_{x\ {\rm near}\ L_2}
\ee
The spectral equation states that  ${\cal S}$  converts
$\ \sin\Big(k(x+L_1)\Big)\sim e^{ik(x+L_1)}-e^{-ik(x+L_1)}\ $ into
$\ \sin\Big(k(x-L_2)\Big) \sim e^{ik(x-L_2)}-e^{-ik(x-L_2)}$.
For the $\delta$-function potential, the scattering matrix is just
\be
{\cal S} = \left(\begin{array}{cc}
	1-\frac{u}{2ik} & -\frac{u}{2ik} \\ \\
	\frac{u}{2ik}& 1+\frac{u}{2ik} \end{array}\right)
\ee
For the two isolated barriers, the scattering matrix is a product
\be
{\cal S}_1 \circ {\cal S}_2 =
{\cal S}_1 \cdot \left(\begin{array}{cc} e^{ikl_{12}} & 0 \\ 0 & e^{-ikl_{12}}\end{array}\right) \cdot {\cal S}_2
\ee
where $l_{12}$ is the distance between the two and so on:
in general one has a product $\circ_i {\cal S}_i$.

Changing the shape of the potential reduces to a composition of reshufflings,
when constituents of the product change order, i.e. to a composition of operations:
\be
{\cal K}_{j}: \ \ \ \circ_{i<j} {\cal S}_i\circ {\cal S}_j\circ {\cal S}_{j+1} \circ_{i>j+1} {\cal S}_i
\ \longrightarrow \
\circ_{i<j} {\cal S}_i\circ {\cal S}_{j+1}\circ {\cal S}_{j} \circ_{i>j+1} {\cal S}_i
\ee
This operation is of course very familiar:
in the theory of quantum groups, if ${\cal S}_i$ are the group elements \cite{MV,MV2}, this permutation
is described by the quantum ${\cal R}$-matrix, i.e. one can suspect that actually
${\cal K}_j = {\cal R}_j$ and, further, ${\cal R}_{j+1} = {\cal U}_j{\cal R}_j{\cal U}_j^\dagger$,
where ${\cal U}$ is the quantum mixing (Racah) matrix, see \cite{MMMII}.
Then from this data it is straightforward to build invariants: these are the ordinary
knot invariants in the Reshetikhin-Turaev-Witten (RTW) formalism (graded traces and certain matrix elements of
the ordered products of the ${\cal R}$-matrices), which in the context of wall-crossing theory
are known as KS invariants.

To make this story really non-trivial, one needs to promote scattering matrices ${\cal S}$ to
operator-valued quantities $\hat {\cal S}$.
In quantum group theory this is achieved by making the elements of the algebra of functions
non-commutative, in quantum mechanics it is enough to introduce internal degrees of freedom
like spins or, more generally, to consider matrix models (e.g. matrix quantum mechanics).
This makes ${\cal R}$- and ${\cal U}$-matrices different from just ordinary permutation matrices:
they start to realize the far less simple braid group structures.

Emergence of braids is, of course, not universal for quantum
mechanical problems, they arise only when the space is
2-dimensional and there are topologically different ways
to adiabatically carry one point around another, producing
a Berry phase.
This is, however, a {\it generic} situation for
algebraically integrable dynamics,
where the separation of variables reduces the study to a
complex curve (sometime called spectral or
Seiberg-Witten curve, the Liouville torus being its Jacobian).
Though formally integrable systems are pretty rare,
there is a growing evidence that typical {\it effective}
field theory obtained after integration over fast variables
is integrable \cite{MV2,UFN3},
and this explains the growing interest to this type of theories.

Even in this integrability context,
naturally appearing representations of the braid group can be quite sophisticated
and difficult to study.
Still there are two immediate classes of examples (in addition to the ordinary Verma modules
of ordinary quantum groups like $SU_q(N)$ widely used in conventional knot theory).
One of them is provided by the WKB limit of quantum mechanics, where the ${\cal R}$-matrices
actually describe reshufflings of the Stokes lines.
Another one is provided by the modular transformations of conformal blocks:
the modular kernels $T_i$ and $S_i$ provide an interesting set of ${\cal R}_i$ and ${\cal U}_i$
matrices, which can be used to construct {\it a priori} new families of knot invariants.
In the simplest case, however, the family is not new: one gets just the ordinary Jones
polynomials (and probably HOMFLY at the next step), but more sophisticated examples
seem capable to provide a long awaited group theory (RTW) interpretation of the Hikami invariants.

The plan of the paper is as follows.

We begin in sec.2 from the general review of the WKB approach
involving the theory of Stokes lines, their reshufflings
and KS invariants.

Then, in sec.3 we consider from this perspective the standard example
of the double-well potential.

After that, in sec.4 we switch to matrix models and reformulate the problem
in terms of the operator-valued (check) resolvents.

In sec.5 we consider KS/RTW invariants, associated with the simplest knots and links
and show that the ${\cal R}$-matrices, provided by the modular transformations
of conformal blocks, give rise to various types of knot invariants:
the Jones polynomials and the Hikami integrals.

A natural part of this presentation are the distinguished (Fock-Goncharov \cite{FG})
coordinates on the moduli space provided by the WKB theory,
where the ${\cal R}$-matrices act via peculiar rational transformations
(known also as {\it mutations} in cluster algebra \cite{CA} and related to discrete change of coordinates
in the algebra of functions \cite{Pop}).

Conclusion in sec.6 describes (incomplete) list of relations
between different subjects in theoretical physics,
which are brought together by consideration of the wall-crossing phenomena.

\section{Wall crossing formulas as a piece of the WKB theory} \label{sec:WKB}

\subsection{Asymptotic behavior}
In Seiberg-Witten (SW) theory describing the low-energy limit of $N=2$ supersymmetric gauge theory \cite{SW},
the central charge and the mass of an excitation
are given by the contour integrals of the SW differential $\lambda$:
\be\label{intrep}
Z_\gamma = \oint_\gamma \lambda, \ \ \ \ \ \
M_\gamma = \oint_\gamma |\lambda|
\ee
BPS states have $M=|Z|$, therefore they  are associated
with the Stokes contours $\sigma(x)\in\Sigma$ on the spectral surface $\Sigma$,
such that the Seiberg-Witten differential $\lambda$ (which is a meromorphic differential on $\Sigma$ whose variations with respect to moduli are holomorphic) along these contours has a definite phase $\phi_\sigma$:
\be\label{BPS}
\hbox{Arg}\left( e^{-i\phi_\sigma}\lambda \Big[\sigma(x)\Big]\right) = 0
\ee
so that
\be
\hbox{Arg}\left(e^{-i\phi_\sigma}\int_{\sigma(x)}\lambda \right)= 0
\ee
If the gauge theory is described via M-theory \cite{WitMth},
then these $\sigma$ are interpreted as intersections of the main $M5$-brane
with the $M2$-branes, see \cite{Mikh}.
The spectral surface $\Sigma$ is a ramified covering of the original bare curve $\Sigma_0$,
and $\lambda$ is the eigenvalue of the Lax 1-form \cite{SWint}.
The mass of the BPS state is given by the absolute value of the same integral,
and the mass is finite when the contour is closed.
To make the Stokes contour closed, one should adjust the phase $\phi_\sigma$,
this possibility to adjust the phase of the Planck constant is the
main new peculiarity of the BPS state counting as compared to the usual
WKB theory.
When the moduli of the spectral surface change, so do the phase and the
shape of the contour,
and at some values of moduli the change can be abrupt: discontinuous.
Such a jump in the multiplicities of the BPS states takes place along the
real-codimension-one surfaces in the moduli space
and is called the wall-crossing phenomenon.
What remains invariant are peculiar combinations of multiplicities,
encoded in the form of Kontsevich-Soibelman formula.

Our purpose in this paper is to discuss a pedestrian approach to
this kind of problems, relating them to the elementary textbook
consideration of Stokes phenomena for the WKB approximation.

For this purpose we consider the Wilson line
\be
W_\Gamma(\phi) = P\exp\left(\frac{ie^{i\phi}}{\hbar} \int_\Gamma {\cal L}\right)
\ee
of the $N\times N$-matrix valued Lax form along an open contour $\Gamma\in \Sigma_0$
on the bare Riemann surface $\Sigma_0$.
The $N$ eigenvalues of ${\cal L}$ define the Seiberg-Witten differential
$\lambda_i$ on the $N$ sheets of the spectral surface $\Sigma$ which
$N$ times covers $\Sigma_0$, so that one can roughly write
$W_\Gamma(\phi)$ as
\be
w_\Gamma(\phi) = {\rm diag}\left\{
\exp\left(\frac{ie^{i\phi}}{\hbar}\int_{\gamma_{i}} \lambda_{i}\right)
\right\}
\ee
where $\gamma_{i}$ are pre-images of $\Gamma$ on $\Sigma$.
However, if one wishes to treat $w_\Gamma$ as a quasiclassical
approximation to an evolution operator
for some quantum-mechanical system,
then one should switch between different branches $i$ when $\Gamma$
crosses the Stokes lines.
\begin{figure}[htbp]
\begin{center}
\includegraphics[scale=0.5]{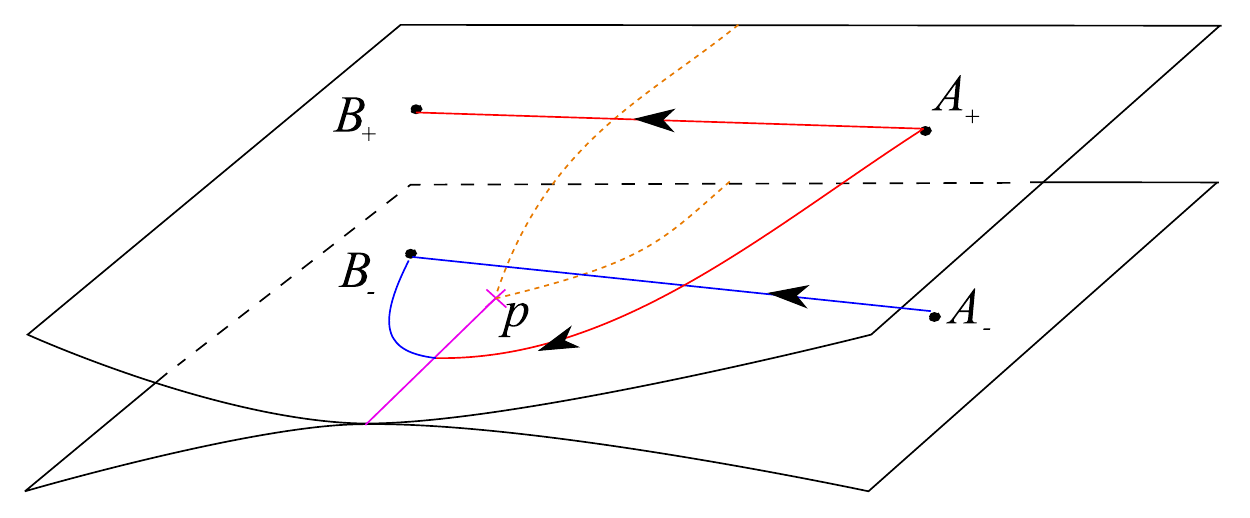}
\caption{Intersecting one Stokes line.\label{fig:1cut}}
\end{center}
\end{figure}
For example, if one has a two-sheet covering with $\lambda_\pm = \pm\lambda$,
and one considers the evolution along the contour, which goes from the point $A$ to $B$
and crosses a Stokes line, originating at ramification point $P$ (see Fig.\ref{fig:1cut}), then
\be
\Psi(B) =
\left(\begin{array}{c}\psi(B_+) \\ \psi(B_-)\end{array} \right)
= \left(\begin{array}{ccc}
\exp \left(\frac{ie^{i\phi}}{\hbar}\int_{A_+}^{B_+} \lambda\right)
& & \exp \left(\frac{ie^{i\phi}}{\hbar}\left(\int^{B_+}_{P}\lambda
- \int^{P}_{A_-}\lambda\right)\right) \\ \\
0 & & \exp \left(-\frac{ie^{i\phi}}{\hbar}\int_{A_-}^{B_-} \lambda\right)
\end{array}\right)
\left(\begin{array}{c}\psi(A_+) \\ \psi(A_-)\end{array} \right)
\ee
while for the inverse path from $B$ to $A$ one would rather encounter a low-triangular matrix.
Thus, instead of the naive $w_\Gamma$ one gets for the quasiclassical Abelization of
the Wilson operator $W_{AB}(\phi)$ a sum of three elementary matrices
\be
\Psi(B) = \Big(\X_{B_+A_+}(\phi) + \X_{B_+PA_-}(\phi)+ \X_{B_-A_-}(\phi)\Big)\Psi(A)
\ee
\begin{figure}[htbp]
\begin{center}
\includegraphics[scale=0.7]{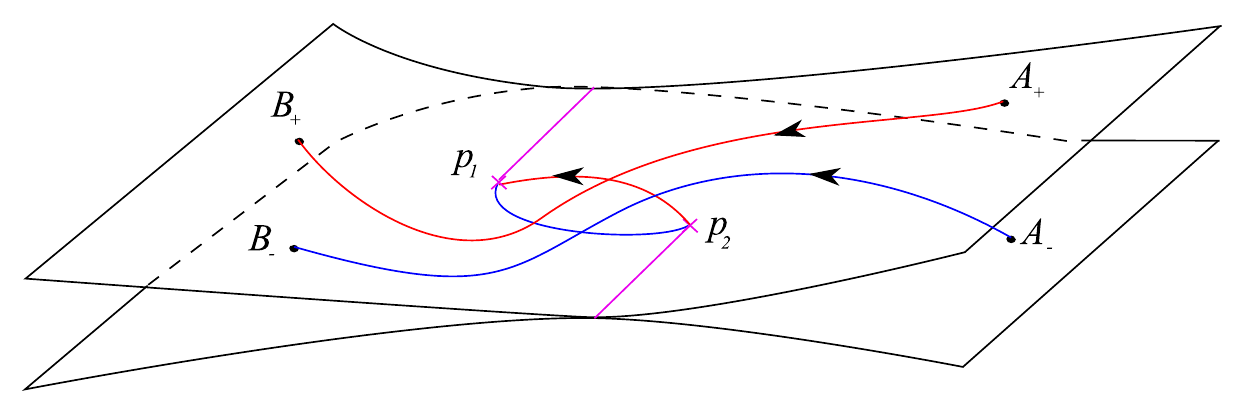}
\caption{Intersecting two Stokes lines.\label{fig:2cut}}
\end{center}
\end{figure}

\subsection{Jumps of WKB network topology on the curve}

Let us call the set of WKB lines (WKB network) $\Gamma\in\Sigma_0$ such that its preimages compose the contour $\sigma$ satisfying (\ref{BPS}).
The WKB network gives a triangulation of the spectral surface $\Sigma_0$. This triangulation depends on the phase $\phi$ and it can jump at some specific critical values $\phi_c$ making a kind of flips of the triangulation.

When a flip occurs, two different Stokes lines merge into a single line of finite length (we denote its pre-image on $\Sigma$ as $\gamma_c$) so that the integral (\ref{intrep}) giving the central charge $Z_{\gamma_c}=\oint\lm_{\gamma_c}\lambda$ becomes convergent. We can immediately define the value of the critical phase as
\be
\phi_c=\mathop{\rm Arg}Z_{\gamma_c}
\ee

Now we observe how the value of the asymptotic expansion of a Wilson operator changes.
If two Stokes lines which originate at the ramification points $P_1$ and $P_2$ were crossed on the way from $A$ to $B$ (see Fig.\ref{fig:2cut}), we would obtain a Wilson operator
\be
\begin{split}
W_{AB}(\phi_-)=\X_{B_+A_+}(\phi_-)+\X_{B_-A_-}(\phi_-) + \X_{B_+P_1A_-}(\phi_-)
+ \X_{B_-P_2A_+}(\phi_-)+\underline{\X_{B_-P_2P_1A_-}(\phi_-)}=\\
=\X_{B_+A_+}(\phi_c)+(1+\X_\sigma)\X_{B_-A_-}(\phi_c) + \X_{B_+P_1A_-}(\phi_c)
+ \X_{B_+P_2A_-}(\phi_c)
\end{split}
\ee
where $\phi_-=\phi_c-0$.
For the other value of phase $\phi_+=\phi_c+0$ configuration of the Stokes lines
can be different, and one obtains {\it another} decomposition
\be
\begin{split}
W_{AB}(\phi_+)=\X_{B_+A_+}(\phi_+) + \X_{B_+P_1A_-}(\phi_+)
+ \X_{B_-P_2A_+}(\phi_+)+\underline{\X_{B_-P_1P_2A_-}(\phi_+)}=\\
=(1+\X_\sigma)\X_{B_+A_+}(\phi_c)+\X_{B_-A_-}(\phi_c) + \X_{B_+P_1A_-}(\phi_c)
+ \X_{B_+P_2A_-}(\phi_c)
\end{split}
\ee
differing not only by the value of $\phi$ but also by reordering
of the points $P_1$ and $P_2$.
At the critical value $\phi_c$ of phase, where reshuffling of the
Stokes lines occurs and a closed Stokes line $\sigma$ appears,
the two expressions differ exactly by $\X_\sigma$, i.e.
an abrupt reshuffling takes place
\be
W_{AB} \ \longrightarrow \ \hat K_\sigma W_{AB}
\ee
where a morphism $\hat K$ acts as
\be\label{flip}
\hat K_\sigma: \ \ \X_\alpha = (1+\X_\sigma)^{<\sigma,\alpha>} \X_\alpha
\ee

Note that this is not an ordinary operator. This morphism acts as a change of coordinates on the moduli space of flat connections. So we apply it to every term in the sum independently, in detail:
\be
\begin{split}
\hat{K}_{\sigma}(W_{AB}(\phi_-))=\hat{K}_{\sigma}(\X_{B_+A_+}(\phi_c))+(1+\hat{K}_{\sigma}(\X_\sigma))\hat{K}_{\sigma}(\X_{B_-A_-}(\phi_c)) + \hat{K}_{\sigma}(\X_{B_+P_1A_-}(\phi_c))
+ \hat{K}_{\sigma}(\X_{B_+P_2A_-}(\phi_c))=\\
=(1+\X_\sigma)^{\underbrace{\footnotesize{\langle\sigma,B_+A_+\rangle}}_{1}}\X_{B_+A_+}(\phi_c)+
\left(1+(1+\X_\sigma)^{\underbrace{\footnotesize{\langle\sigma,\sigma\rangle}}_{0}}\X_{\sigma}\right)(1+\X_\sigma)^{\underbrace{\footnotesize{\langle\sigma,B_-A_-\rangle}}_{-1}}\X_{B_-A_-}(\phi_c)+\\
+(1+\X_\sigma)^{\underbrace{\footnotesize{\langle\sigma,B_+P_1A_-\rangle}}_{0}} \X_{B_+P_1A_-}(\phi_c)
+ (1+\X_\sigma)^{\underbrace{\footnotesize{\langle\sigma,B_+P_2A_-\rangle}}_{0}}\X_{B_+P_2A_-}(\phi_c)=W_{AB}(\phi_+)
\end{split}
\ee

\subsection{Non-trivial moduli space invariants: wall-crossing formulae in the moduli space}

As we have seen the asymptotics of Wilson lines is {\bf not} smooth. The discontinuity is of order $w_{\sigma}(\phi_c)$, which is asymptotically small.

Nevertheless, the very value of the observable is expected to be smooth. This fact allows one to construct non-trivial invariants of the morphisms $\hat K$ on the Coulomb branch, called spectrum generators \cite{GMN} tightly related to the spectra of BPS states arising in the effective theory.

In general, when one starts increasing the phase $\phi$ from $0$ to $\pi$,
a number of reshufflings take place, when particular closed Stokes lines
$\sigma_a$ appear at critical values $\phi_a$, and disappear with the
further increase of $\phi$.
This provides a sequence of actions
\be
\overleftarrow{\prod_a} \hat K_{\sigma_a}
\ee
The number of factors here is actually the number of BPS states
on the given spectral curve, i.e. at the given point of the
moduli space.
If one now starts changing {\it moduli} of the spectral curve
this very product can change, reflecting the change of the
ordered set ${\cal A}$ of the BPS states, including their number
(the number of factors in the product),
and the order in which they occur with increase
of the phase $\phi$.
However, at the domain wall in moduli space (at the hypersurface
of marginal stability) given by the condition $\phi=\phi_c$ the two different products should coincide:
\be
\boxed{
\overleftarrow{\prod_{a\in {\cal A}}} \hat K_{\sigma_a} =
\overleftarrow{\prod_{b\in{\cal B}}} \hat K_{\sigma_b}
}
\ee
thus we obtain the Kontsevich-Soibelman (KS) invariant, taking values in functors,
acting on the space of $\X$-variables often called the
Fock-Goncharov coordinates of the flat connection moduli space.

\paragraph{Basic example:} For two conjugated $A$ and $B$ cycles on a torus
with $<A,B>=1$ the KS relation states:
\be
\hat K_{A}\hat K_B= \hat K_B\hat K_{A+B}\hat K_A
\label{KK=KKK}
\ee
where the operator action is defined as
\be
\hat K_{mA+nB} \X_\gamma = (1+\X_A^m\X_B^n)^{m<A,\gamma>+n<B,\gamma>}\X_\gamma
\ee
Note that the coordinates $\X_A\X_B=\X_B\X_A$ commute, and also
$\X_{A+B} = \X_A\X_B$, while neither of these is true for the operators $\hat K$.
With these definitions eq.(\ref{KK=KKK}) is just an identity, indeed,
applying both sides to, say $\X_A$, one gets
\be
\hat K_{A}\hat K_B \X_A = \hat K_A \frac{1}{1+\X_B}\X_A = \frac{1}{1+(1+\X_A)\X_B}\X_A
= \frac{\X_A}{1+\X_B+\X_A\X_B},
\ee
and
\be
\hat K_B\hat K_{A+B}\hat K_A\X_A = \hat K_B\hat K_{A+B}\X_A =
 \hat K_B \frac{1}{1+\X_A\X_B}\X_A
= \frac{1}{1+\frac{1}{1+\X_B}\X_A\X_B}\cdot \frac{1}{1+\X_B}\X_A =
\frac{\X_A}{1+\X_B+\X_A\X_B}
\ee
Similarly, in application to $\X_B$:
\be
\hat K_{A}\hat K_B \X_B = (1+\X_A)\X_B,
\ee
and
\be
\hat K_B\hat K_{A+B}\hat K_A\X_B = \hat K_B\hat K_{A+B} (1+\X_A)\X_B
= \hat K_B \left(1+\frac{1}{1+\X_A\X_B}\X_A\right)(1+\X_A\X_B)\X_B = \nn\\
= \left(1+\frac{1}{1+\frac{1}{1+\X_B}\X_A\X_B}\frac{1}{1+\X_B}\X_A\right)
\left(1+\frac{1}{1+\X_B}\X_A\X_B\right)\X_B =
\frac{1+\X_A+\X_B+\X_A\X_B}{1+\X_B}\X_B = (1+\X_A)\X_B\nn
\ee

\bigskip

\section{Classical problem of quantum mechanics: double well potential}

Consider the Schr\"odinger equation with the quartic potential
\be
\left[\hbar^2 \p_z^2 -(z-x_1)(z-x_2)(z-x_3)(z-x_4)\right]\Psi(z)=0
\ee
Depending on the choice of zeroes $x_k$, the structure of levels is rather different.

At the first glance, this may seem a little bit controversial. According to the well-known theorem in ODE theory, solutions to the \emph{Cauchy} problem are continuous functions of parameters if the coefficients of equation are continuous. Nevertheless, the problem of finding eigenvalues of \emph{self-adjoint} operators (\emph{Sturm-Liouville} problem) is quite different. Once found integrable in the usual Hilbert norm, the eigenfunctions at some chosen values of parameters are not expected to keep integrability at another choice of the parameters.

The integrability of function depends on its asymptotic behavior. In this particular case there are two asymptotics $e^{\pm \frac{z^3}{\hbar}}$. Choose two linearly independent solutions $\Psi_{1,2}$ with the asymptotics behaviour being
\be
\Psi_1(z)\mathop{\sim}_{z\rightarrow\pm\infty}c_{1+}(\pm\infty)e^{\frac{z^3}{\hbar}}+c_{1-}(\pm\infty)e^{-\frac{z^3}{\hbar}}\\
\Psi_2(z)\mathop{\sim}_{z\rightarrow\pm\infty}c_{2+}(\pm\infty)e^{\frac{z^3}{\hbar}}+c_{2-}(\pm\infty)e^{-\frac{z^3}{\hbar}}
\ee
We define an ``${\cal S}$-matrix'' as
\be
{\cal S}=\left(\begin{array}{cc}
\sigma_{++} & \sigma_{-+} \\
\sigma_{+-} & \sigma_{--}
\end{array}\right)=
\left(\begin{array}{cc}
c_{1+}(+\infty) & c_{1-}(+\infty) \\
c_{2+}(+\infty) & c_{2-}(+\infty)
\end{array}\right)^{-1} \left(\begin{array}{cc}
c_{1+}(-\infty) & c_{1-}(-\infty) \\
c_{2+}(-\infty) & c_{2-}(-\infty)
\end{array}\right)
\ee
Note that this matrix is independent of the choice of the basis in solutions. For a generic choice of parameters this ${\cal S}$-matrix is unphysical.

To define an eigenfunction we require it to be integrable for real $\hbar$, i.e. to behave as
\be
\Psi(z)\mathop{\sim}_{z\rightarrow\pm\infty}e^{\mp \frac{z^3}{\hbar}}
\ee
This imposes a condition on the ${\cal S}$-matrix entries
\be
\sigma_{--}(x_1,x_2,x_3,x_4)=\sigma_{++}(x_1,x_2,x_3,x_4)=0
\ee

The crucial point is that the ${\cal S}$-matrix is \emph{discontinuous} on the moduli space $(x_k,\hbar)$ (see also \cite{Rub}).

To observe this, we consider two well-known physical situations:
\begin{itemize}
\item[I.] The ground energy level is below the level of the wall between the wells: all the zeroes $x_k$ are real. The problem can be described by a particle localized either at the left well or at the right well so that there are almost degenerate two levels (with the wavefunctions symmetric and antisymmetric w.r.t. interchanging the wells) which differ only due to
    instanton jumps between the wells.

Topology of the WKB lines is depicted in Fig.\ref{fig:below}

\begin{figure}[htbp]
\begin{center}
\includegraphics[scale=0.6]{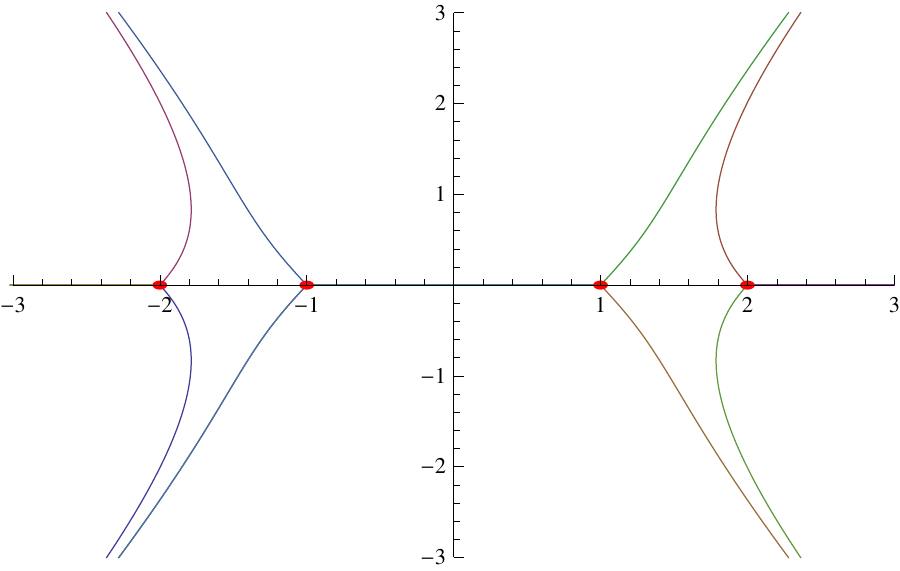}
\caption{Topology of the WKB lines when all zeroes are real \label{fig:below}}
\end{center}
\end{figure}

The first WKB approximation gives the following expression for $\sigma_{--}$:
\be
\sigma_{--}(x_k)\sim-4 \underbrace{\sinh\left(\frac{1}{\hbar}\int\lm_{x_1}^{x_2}\lambda\right)}_{\mbox{first well level}}\underbrace{\sinh\left(\frac{1}{\hbar}\int\lm_{x_3}^{x_4}\lambda\right)}_{\mbox{second well level}}+\underbrace{e^{-\frac{2}{\hbar}\int\lm_{x_2}^{x_3}\lambda}}_{\mbox{instanton}}e^{-\frac{1}{\hbar}\int\lm_{x_1}^{x_2}\lambda}e^{-\frac{1}{\hbar}\int\lm_{x_3}^{x_4}\lambda}
\ee
\item[II.] The ground energy level is above the level of the wall between the wells: two zeroes are real, two zeroes have opposite imaginary parts. In this case there are no two almost degenerate energy levels.

Topology of the WKB lines is depicted in Fig.\ref{fig:above}

\begin{figure}[htbp]
\begin{center}
\includegraphics[scale=0.6]{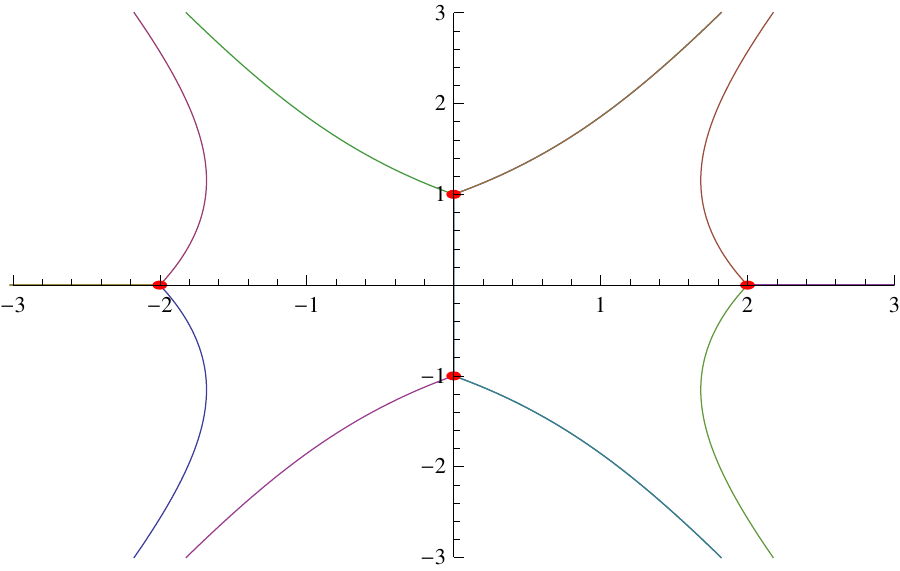}
\caption{Topology of the WKB lines when two zeroes are real and two zeroes have opposite imaginary parts\label{fig:above}}
\end{center}
\end{figure}

The first WKB approximation gives the following expression for $\sigma_{--}$:
\be
\sigma_{--}(x_k)\sim -e^{\frac{1}{2\hbar}\oint\lm_{\infty}\lambda}-2\sin\left(\frac{1}{\hbar}\int\lm_{x_1}^{x_4}\lambda\right)
\ee

\end{itemize}

${\cal S}$ jumps discontinuously, the jump being described by the operator $\hat K$ which is a counterpart of the KS-operator in ordinary quantum mechanics:
\be
\hat K({\cal S}^{\rm (I)})={\cal S}^{\rm (II)}
\ee

However, ${\cal S}$ depends only on the point $x_k$ \emph{not belonging to the path} in the moduli space. Hence, all the jumps along a closed contractible loop should cancel
\be
\prod\lm_{\rm loop }\hat K=1
\ee

\section{Check- $=$ ``quantum, refined'' }
\subsection{Intuitive remarks}
In many applications matrix elements of the $N\times N$
Lax form are themselves operators. As we saw, such are the
Seiberg-Witten differentials $\lambda^{(i)}$ and the
Abelianized monodromies $w_\Gamma$.
In such cases, the Kontsevich-Soibelman relations include
the Fock-Goncharov coordinates,
which take values in the operators
rather than in numbers.

Here first we give some intuitive remarks about possible ``refinement'' of the Abelianization map construction then we give a more solid description using ``check''-operators \cite{AMM1,AMM2,AMM3}.

Speculations are quite ``hand-waving'' so far, nevertheless we would try to draw several important conclusions:
\begin{enumerate}
\item \emph{Gauge covariance}: The commutation relations inherited from the natural Poisson structure on the Lax operator \cite{GMN2}
 \be\label{LaxPois}
\varpi=\int d^2 z\;\mathop{\rm Tr\;} \delta {\cal L}\wedge \overline{\delta {\cal L}}
\ee
is not gauge invariant. Similarly, the same-time commutator in Chern-Simons theory is known to be induced by (\ref{LaxPois}) only in the temporal gauge.
    There is no problem to define an \emph{invariant} commutator
\be\label{1c}
\left[\oint\lm_{\gamma}\hat\lambda,\oint\lm_{\gamma'}\hat\lambda\right]=\hbar'\left\langle\gamma,\gamma'\right\rangle
\ee
where $\gamma$'s are paths on the spectral curve, and $\langle\star,\star\rangle$ is the cycle pairing on the curve.
\item \emph{``Anomaly''}: The commutator before integration $\left[\hat\lambda^{(i)}(z),\hat\lambda^{(j)}(z')\right]$ is a kind of ``anomalous'': one can smoothly modify the paths unless one touches the intersection points where the commutator contributes. This breaks the initial holomorphicity of the problem, so that the expression $\oint\lm_\gamma\hat\lambda$ depends now on the regular homotopy class of $\gamma$ rather than on the homology class of $\gamma$. Suppose we can choose a representative $[\gamma]$ in the homology class of $\gamma$ without self-intersections, then
\be
\Pexp \oint\lm_{\gamma}\hat \lambda=q^{\wt \gamma}e^{\oint\lm_{[\gamma]}\hat\lambda}
\ee
where $\wt$ is a \emph{writhe}, a signed sum over self-intersections.

Similarly, one defines the coordinates depending only on the homology class $[\gamma]$
\be
\X_{[\gamma]}=\exp\left(\frac{ie^{i\phi}}{\hbar}\int_{[\gamma]} \lambda\right)
\ee
They form a non-commutative algebra
\be\label{algebra}
\X_{[\gamma]}\X_{[\gamma']}=q^{\langle[\gamma],[\gamma']\rangle}\X_{[\gamma]+[\gamma']}
\ee
where $q=\exp\left(ie^{i\phi} \hbar'/2\hbar\right)$ (note that $\hbar$ and $\hbar'$ are generally two independent constants).
To derive this consider a product of two exponents:
\be
\Pexp \oint\lm_{\gamma}\hat \lambda \;\Pexp \oint\lm_{\gamma'}\hat \lambda=\Pexp \oint\lm_{\gamma\circ \gamma'}\hat \lambda
\ee
Where $\circ$ denotes a consequent concatenation of two paths. Equivalently, this relation can be rewritten in terms of $w$-variables:
\be
q^{\wt \gamma} w_{[\gamma]}\; q^{\wt \gamma'} w_{[\gamma']}=q^{\wt(\gamma\circ\gamma')}w_{[\gamma\circ\gamma']}
\ee
Using relations
\be
[\gamma\circ\gamma']=[\gamma]+[\gamma']\\
\wt (\gamma\circ\gamma')=\wt \gamma+\wt \gamma'+ \langle [\gamma],[\gamma']\rangle
\ee
one reproduces algebraic relation (\ref{algebra}). The second relation says that the writhe function is a quadratic refinement of the intersection form, for details see \cite[Appendix C]{GLM}.

As in the previous section, the Wilson lines are polynomials in $\X$-variables, though now over $\IZ[q,q^{-1}]$
\be
\Tr \Pexp\oint L\sim\sum\lm_{\gamma} q^{\wt\gamma}w_{[\gamma]}\label{asymp}
\ee
\end{enumerate}

To conclude this section we mention that this quite heuristic consideration can be applied to physical problems \cite{GLM}. $\hat K$-jumps of expansion (\ref{asymp}) similarly to the jumps discussed in the previous section allow one to calculate characteristics of the BPS spectra in ${\cal N}=2$ SYM theories. These invariants are refined now with a deformation parameter $q$ and take into account the spin of BPS multiplets.

\subsection{Beta-ensemble construction}

Beta-ensembles naturally extend matrix models and inherit their basic properties. The model is given by the partition function of 2d Coulomb gas (here we consider an example when the gas is placed on the sphere)
\be\label{pf}
Z=\prod\lm_i \oint\lm_{\gamma_i} dz_i \prod\lm_{i<j}\left(z_i - z_j\right)^{2\beta}e^{-\frac{1}{g}\sum\lm_i V(T|z_i)}
\ee
where $g$, $\beta$ are two parameters similar to $\hbar$ and $\hbar'$, and the potential $V(T|z)$ determines the moduli space of the partition function: it is parameterized by the parameters $T_k$ of the potential and by the choice of the integration contours. For the sake of definiteness, we choose the potential to be a polynomial and $T_k$ to be coefficients of this polynomial:
\be
V(z)=\sum\lm_{k=0}^n T_k z^k
\ee

We consider only closed contours so that changing the variables
\be\label{variation}
z_i \rightarrow z_i+\frac{\epsilon}{\zeta-z_i}
\ee
does not change the integral, which leads to the following Ward identity in the first order of $\epsilon$
\be
\left\langle\sum\lm_i \frac{1}{(\zeta-z_i)^2}+\beta\sum\lm_{i\neq j}\frac{1}{(\zeta-z_i)(\zeta-z_j)}-\frac{1}{g}\sum\lm_i V'(z_i)\right\rangle =0
\ee

After some algebra this equation can be represented in the following form
\be\label{leq}
(\beta-1)\p_{\zeta}\left\langle\sum\lm_i \frac{1}{\zeta-z_i}\right\rangle+\beta\left\langle\left(\sum\lm_i \frac{1}{\zeta-z_i}\right)^2\right\rangle-\frac{V'(\zeta)}{g}\left\langle\sum\lm_i \frac{1}{\zeta-z_i}\right\rangle+\frac{1}{g}\left\langle\sum\lm_i\frac{V'(\zeta)-V'(z_i)}{\zeta-z_i}\right\rangle=0
\ee
Let us define the resolvent as
\be\label{res}
\rho(\zeta):={1\over Z}\hat{\nabla}(\zeta)Z:=g\sqrt{\beta}\left\langle\sum\lm_i \frac{1}{\zeta-z_i}\right\rangle
\ee
where the operator $\hat{\nabla}(\zeta)$ can be described by the action of $g\sqrt{\beta}\sum {1\over z^{k+1}}{\partial\over\partial t_k}$ on the
partition function with the modified potential $V(z)\rightarrow V(z)-g\sum_kt_kz^k$, treated as a formal (perturbative) series in the variables $t_k$:
\be\label{fpf}
{\cal Z}=\prod\lm_i \oint\lm_{\gamma_i} dz_i \prod\lm_{i<j}\left(z_i - z_j\right)^{2\beta}e^{-\frac{1}{g}\sum\lm_i V(T|z_i)+\sum_{i,k}t_kz_i^k}
\ee
We use $Z$ for the partition function ${\cal Z}$ restricted to all $t_k=0$.

The last term in (\ref{leq}) can be reproduced by the action of the differential operator in $T_k$'s:
\be
\check P(\zeta)=-g^2\sum_{k=2}^n k T_k \sum\lm_{n=0}^{k-2}\zeta^n \p_{ T_{k-2-n}}
\ee
We call such operators {\it check}-operators since they act on moduli $T_k$'s (in variance with $\hat{\nabla}(\zeta)$).
Note that the Miwa transform of $T_k$-moduli
\be
T_k=\sum\lm_j \frac{\alpha_j \lambda_j^k}{k}
\ee
transforms the check-operator $\check P(\zeta)$ into
\be
\check P(\zeta)\sim \sum\lm_j \frac{\p_{\lambda_j}}{\zeta-\lambda_j}
\ee

One also can introduce the check-operator that generates the resolvents:
\be\label{chop}
\check\nabla {\cal Z}=\hat\nabla {\cal Z}
\ee
This operator can be constructed recursively from $\check y:=\sqrt{V'(T|z)^2-4\beta\check P}$, its derivatives and $V'(T|z)$ \cite{AMM2}. The recursion is provided by the $g^2$-expansion, and, in the leading order, $\check\nabla^{(0)} = g\sqrt{\beta}\check y$.

Then the Ward identity can be rewritten in the form
\be\label{WI}
\left[gQ \p_{\zeta}\check{\nabla}(\zeta)+{\bf :}\check{\nabla}(\zeta)^2{\bf :}-{V'(\zeta)\over\sqrt{\beta}}\check{\nabla}(\zeta)+\check P(\zeta)\right]Z=0
\ee
where $Q:=\sqrt{\beta}-{1\over\sqrt{\beta}}$ and the normal ordering means that the operator $\check\nabla(\zeta)$ acts only on ${Z}$, but not on itself.

\subsection{Determinant check-operator: quantizing the spectral curve}\label{sec:quantizingSC}

In the leading order of the WKB approximation (i.e. $g^2\to 0$), the Ward identity (\ref{WI}) becomes the algebraic equation
for the resolvent (\ref{res}):
\be\label{sc1}
\rho^{(0)}(\zeta)^2-\frac{V'(\zeta)}{\sqrt{\beta}}\rho^{(0)}(\zeta)+f(\zeta)=0
\ee
where the polynomial $f(\zeta):=\check P \log Z$. This algebraic equation is nothing but the spectral curve.
Note that the monodromies of these check-operators along $A$- and $B$-periods of these spectral curve form the Heisenberg algebra \cite{AMM2}:
\be
\left[\oint_{A_i}\check\nabla(\zeta)d\zeta,\oint_{B_j}\check\nabla(\zeta)d\zeta\right]=\delta_{ij}
\ee

One may ask what are the ways of quantizing the spectral curve (\ref{sc1}) (which has to become the Baxter equation after quantization). There are two possibilities. One of the possibilities is to consider the limit of (\ref{pf}) when $g^2/\beta=\hbar'^2\to 0$ with $g^2\beta=\hbar^2$ kept fixed. In this limit (called Nekrasov-Shatashvili limit \cite{NS}) the Ward identity (\ref{WI}) gets the Ricatti (or Schr\"odinger) equation equivalent to the Baxter equation \cite{MMS} and corresponding to a quantum integrable system \cite{NS,BS,surop,NRS,Krefl}.

However, this system depends only on one parameter $g\sqrt{\beta}=\hbar$. However, there is a possibility of quantizing the spectral curve which still preserves the $\beta$-ensemble representation for the wavefunction and depends on two parameters $\beta$, $g$. To this end, in \cite{surop} it was suggested to consider the equation for the $\beta$-ensemble average of the would-be determinant in matrix model: the average $\Psi(\zeta)=\left\langle\prod_i(\zeta-z_i)\right\rangle$. To dealing with this average, one introduces another check-operator:
\be
{1\over {\cal Z}}\check {\cal D}_{[1]}(\zeta){\cal Z}:=\left\langle\prod_i(\zeta-z_i)\right\rangle
\ee

In order to understand the meaning of $\Psi(\zeta)$ let us rewrite it as (the number of integrations in the $\beta$-ensemble partition function is denoted through $N$)
\be
\Psi(\zeta)=\zeta^N\left\langle\prod_i\left(1-{z_i\over\zeta}\right)\right\rangle  =
\zeta^N\left\langle \exp\left\{\sum_i\left(1-{z_i\over\zeta}\right)\right\}\right\rangle  =
\zeta^N \left\langle\exp\left( -\sum_k \frac{\sum_i z_i^k}{k\zeta^k}\right)\right\rangle
\ee
This expression is equal both to
\be\label{det1}
\Psi(\zeta)= \zeta^N\frac{1}{{\cal Z}} \exp\left(\int^\zeta dz \sum_k \frac{1}{z^{k+1}}\frac{\p}{\p t_k}\right) {\cal Z}=\frac{
\exp\left({1\over\hbar}\int^\zeta dz\hat\nabla(z)\right) {\cal Z} }{{\cal Z}}
\ee
where $N$ is treated as zero time $t_0$,
and to
\be\label{BA}
 \Psi(\zeta)=\frac{ {\cal Z}\left(t_k-\frac{1}{k\zeta^k}\right)}{{\cal Z}(t_k)}
\ee
Let us consider the case of $\beta=1$ when the $\beta$-ensemble reduces to the Hermitean matrix model. Within the AGT, this case corresponds to the conformal field theory with central charge 1 \cite{AGT,AGT1}. In this case, ${\cal Z}$ is a $\tau$-function of the Toda-chain hierarchy \cite{GMMMO} in time variables $t_k$, and $\Psi(\zeta)$ given by formula (\ref{BA}) is the Baker-Akhiezer function. It corresponds to insertion of the fermion $\psi(\zeta)$ at the point $\zeta$ (and a fermion $\psi$ at infinity). This can be easily understood, since in $c=1$ theory (free fields) the fermion is described in terms of the free field $\phi(\zeta)$ by the exponential $:\exp(i\phi(\zeta)):$ which being inserted into the conformal correlator representation of the matrix model \cite{MMM} gives exactly the determinant \cite{surop}. This fermion describes the simplest fundamental representation of the $SL(N)$ group which can be understood from the realization of its $N$-plet as \cite{GKLMM}
\be
\psi_i|0>=T_-^{i-1}|0>
\ee
where the fermion modes are defined by $\psi(\zeta)=\sum_i^N\psi_i\zeta^i$ and $T_-=\sum_i^{N-1}T_{-\alpha_i}$ is the sum of all raising operators associated with the negative simple roots of $SL(N)$. Then all other fundamental representations are defined as products of fermions. This connection with the fundamental representations can be also made manifest via the expansion of the determinant into the fundamental representations:
\be
\prod_i\left(1-{z_i\over\zeta}\right)=\sum_k\chi_{[1^k]}(z_i)\left(-{1\over\zeta}\right)^k
\ee
where $\chi_{[1^k]}(z_i)$ is the Schur function, i.e. the characters of group $SL(N)$ associated with the fundamental representations.

One now can convert the Ward identity (\ref{WI}) into the equation for $\Psi(\zeta)$. To this end, one needs to rewrite
the Ward identity in terms of shifted time-variables and of operators acting on ${\cal Z}$:
\be
\left[gQ \p_{\zeta}\hat{\nabla}(\zeta)+\hat{\nabla}(\zeta)^2-{1\over\sqrt{\beta}}\left(V'\Big(t_k-\frac{1}{k\zeta^k}\Big|\zeta\Big)\hat\nabla(\zeta)\right)_-\right]
{\cal Z}\left(t_k-\frac{1}{k\zeta^k}\right) = 0
\ee
where the subscript "-" refers to negative powers of $\zeta$ and $\hat\nabla(\zeta)$ does not need to be normal ordered, since it does not act on itself. We will use further
\be
\left(V'\Big(t_k-\frac{1}{k\zeta^k}\Big|\zeta\Big)\hat\nabla(\zeta)\right)_- =
g\sqrt{\beta}\sum_{n\geq -1} \frac{1}{\zeta^{n+2}} \sum_{k\geq 1}
k\left(t_k-\frac{1}{k\zeta^k}\right)\frac{\p}{\p t_{k+n}}
= \nn \\
= \Big(V'(t|\zeta)\hat\nabla(\zeta)\Big)_-
- g\sqrt{\beta}\sum_{k,n} \frac{1}{\zeta^{k+n+2}}\frac{\p}{\p t_{k+n}}
= \Big(V'(t|\zeta)\hat\nabla(\zeta)\Big)_- - \p_\zeta\hat\nabla(\zeta)
\ee
where at the last step we compared the coefficient $\ k+1\ $ in
$\p_\zeta\hat\nabla(\zeta) =g\sqrt{\beta} \sum_{m\geq 0} \frac{m+1}{z^{m+2}}\frac{\p}{\p t_m}$
with $\sum_{\stackrel{k\geq 1}{n\geq -1}} \delta_{k+n, m} = m+1$.

Since, for the ``non-normalized'' $\tilde\Psi(\zeta):={\cal Z}\left(t_k-\frac{1}{k\zeta^k}\right)$, one directly checks that  $g\sqrt{\beta}\p_\zeta\tilde\Psi(\zeta)= \hat\nabla(\zeta)\tilde\Psi(\zeta)$ and $g^2\beta\p_\zeta^2\tilde\Psi =
\Big(\hat\nabla^2(\zeta) + g\sqrt{\beta}\p_\zeta\hat\nabla(\zeta)\Big)\tilde\Psi(\zeta)$, one finally obtains
a \emph{differential} equation
\be
\boxed{
\left[g^2\beta\p_{\zeta}^2-V'(\zeta)\p_{\zeta}+\check P(\zeta, T_k)\right]\Psi(\zeta)=0}
\ee
which is a quantization of the algebraic spectral curve which depends on the both parameters of deformation $g$ and $\beta$. With rescaling the wave function $\Psi(\zeta)=\exp\Big(V(\zeta)/(g^2\beta)\Big)\tilde\Psi(\zeta)$, one can recast this equation to the form
\be\label{keq}
\left[g^2\beta\p_{\zeta}^2+{1\over 2}\Big(V''(\zeta)-{V'^2(\zeta)\over 2g^2\beta}+{1\over g^2\beta}\Big[\check P(\zeta, T_k)V(\zeta)\Big]\Big)+\check P(\zeta, T_k)\right]\tilde\Psi(\zeta)=0
\ee

Now one can easily construct action of the determinant check operator on the partition function using (\ref{det1}):
\be
\left\langle\prod_i(\zeta-z_i)\right\rangle {\cal Z}= e^{\frac{1}{\hbar}\int\lm^{\zeta}dz\hat{\nabla}(z)} {\cal Z}=
 {\bf :}e^{\frac{1}{\hbar}\int\lm^{\zeta}dz\check{\nabla}(z)}{\bf :} {\cal Z}
\ee

It is certainly clear how to construct the ordinary operator itself in terms of time variables $t_k$ \cite{AMM45}:
\be
\hat{\cal D}_{[1]}(\zeta)={\bf :}\exp\left(\int\lm^{\zeta}dz\hat J(z)\right){\bf :}\\
\hat J(z)=\sum_k \Big({1\over 2}kt_kz^{k-1}+{1\over z^{k+1}}{\partial\over\partial t_k}\Big)
\ee
since this is nothing but the exponential of the free field realized via its action on functions of time variables $t_k$ which corresponds to the fermion (see above). The normal ordering here means that all the $t_k$-derivatives are put to the right.

The equation for the determinant operator is of the second order in the variable $\zeta$, and as we move along some closed contour $\gamma$ the two solutions might have some monodromy. We define a gauge-invariant operator $\check{\cal O}_{[1]}$ as a trace of this monodromy matrix
\be
\check{\cal O}_{[1]}(\gamma)=\Tr \mathop{\rm Mon}(\gamma,\zeta) \check{\cal D}_{[1]}(\zeta)
\ee
If one takes unto account the both branches and corrections from ``the measure anomaly'', extra conjugation factors \cite{GMM}:
\be
\boxed{\check{\cal O}_{[1]}(\gamma)=\sum\lm_{r=\pm} Z^{-1}(V\rightarrow 0,\check{\nabla}_{(r)}) e^{\frac{1}{g\sqrt{\beta}}\oint\lm_{\gamma}dx\check{\nabla}_{(r)}(x)}Z(V\rightarrow 0,\check{\nabla}_{(r)})}
\ee

\bigskip

Hence we establish the following dictionary between integrable models and beta-ensembles:
\be
\begin{array}{ccc}
\lambda &\leftrightsquigarrow& \check{\nabla}\\
\hbar &\leftrightsquigarrow& g\sqrt{\beta} \\
\hbar' &\leftrightsquigarrow& g/\sqrt{\beta}\\
\Tr_R \Pexp\oint_\gamma {\cal L} &\leftrightsquigarrow& \check{\cal O}_{R}(\gamma)\\
w_{\gamma} &\leftrightsquigarrow& \exp\Big(\frac{1}{g\sqrt{\beta}}\oint\lm_{\gamma}\check{\nabla}\Big)
\end{array}
\ee

\subsection{Higher weight operators and spectral covers}
We denoted the determinant operator by a subscript $[1]$ to stress that the operator defined in this way represents a Wilson line in the fundamental representation. Hence, one expects a natural generalization
\be
\Tr_R \Pexp \oint_{\gamma} L \leftrightsquigarrow  \check{\cal O}_R(\gamma)
\ee
A naive expression for this operator is expected to be the trace of monodromy of the determinant operator $\check{\cal D}_{R}(\zeta)$ that inserts something like $\det_R(\zeta-M)$ into the beta-ensemble averaging:
\be
\check{\cal O}_{R}(\gamma)=\Tr \mathop{\rm Mon}(\gamma,\zeta) \check{\cal D}_{R}(\zeta)
\ee
We expect that these operators should satisfy the Wilson loop OPE algebra
\be
\check{\cal O}_R\otimes \check{\cal O}_{R'}=\sum\lm_{Q\vdash |R|+|R'|}C_{R,R'}^Q\check{\cal O}_Q
\ee
where $C_{R,R'}^Q$ are the corresponding Clebsh-Gordon coefficients.

In the case of $SU(2)$ these operators are usually associated to degenerate operators in the Liouville conformal field theory (see s.5.2 below)
\be
\check{\cal D}_{[j]}(\zeta) \sim \Phi_{(j+1,1)}(\zeta)
\ee
with the following fusion algebra
\be
\Phi_{(j+1,1)} \otimes \Phi_{(j'+1,1)}=\bigoplus\lm_{s=|j-j'|}^{j+j'}\Phi_{(s+1,1)}
\ee
Translating this remark back to the beta-ensemble framework we define
\be
\check{\cal D}_{[j]}(\zeta)Z:=\left\langle\prod\lm_i (\zeta-z_i)^j\right\rangle Z
\ee

Nevertheless the naive form of $R$-dependence reveals itself in the form of the spectral curve. One can define a generic spectral cover as
\be
\mathop{\rm Det}\nolimits_R(\lambda-L(z))=0
\ee
According to this remark, for a symmetric representation $[r]$ one expects the spectral curve to be a polynomial in $\lambda$ of degree $r+1$, the same is the order of the differential equation.

Here we present an explicit example of the differential equation satisfied by $\check D_{[2]}$.

We are looking for a variation such that the variation of measure in the partition function (\ref{pf}) can be rewritten as a derivative acting on $\check{\cal D}_{[2]}(\zeta)$. Denote variation (\ref{variation}) as $\delta(\zeta)$, we expect some linear combination of variations $\delta(\zeta)\p_{\zeta}$ and $\p_{\zeta}\delta(\zeta)$ to give a desired result. Indeed,
\be
\begin{split}
\left\{\left(1+\frac{2}{\beta}\right)\delta(\zeta)\p_{\zeta}+\left(1-\frac{2}{\beta}\right)\p_{\zeta}\delta(\zeta)\right\}\check{\cal D}_{[2]}(\zeta)Z=\\
&\left\{\frac{\beta}{2}\p_{\zeta}^3+\frac{1}{g^2}\check T_{[2]}(\zeta)\right\}\check{\cal D}_{[2]}(\zeta)Z=0
\end{split}
\ee
Where $\check T_{[2]}(\zeta)$ is a contribution from the potential $V$:
\be
\begin{split}
\check T_{[2]}(\zeta)=\left(1-\frac{2}{\beta}\right)\left(-\p_{\zeta}\check P(\zeta)+\frac{V'(\zeta)}{2}\p_{\zeta}^2\right)-\left(1+\frac{2}{\beta}\right)\check{P} (\zeta)\p_{\zeta}
+\frac{2}{\beta}V'(\zeta)\left(\frac{1+\beta}{2}\p_{\zeta}^2+\frac{1}{g^2}\check{P}(\zeta)+\frac{V'(\zeta)}{2g^2}\p_{\zeta}\right)
\end{split}
\ee

The operators ${\cal O}_R$ are counterparts of the linear group (Schur) characters for the corresponding representations $R$.

To clarify this point, first assume that the flat connection $A$ is not quantized, then the naive asymptotic form reads
\be
\Tr_R \Pexp \frac{1}{\hbar}\oint\lm_{\gamma}A\sim P_R\left(e^{\frac{1}{\hbar}\oint\lm_{\gamma}\lambda^{(i)}}\right)
\ee
where $\lambda^{(i)}$ are solutions to the equation
\be
\mathop{\rm det}\nolimits_R(\lambda-A)=0
\ee
and the polynomial $P_R$ depends only on the representation $R$. One can make use of this latter fact and assume that $A$ is a constant field. Hence, one can substitute the ordered exponential by the ordinary exponential and omit the integral rewriting this equation as
\be
\Tr_R e^{\frac{\alpha}{\hbar}A}=P_R(e^{\frac{\alpha}{\hbar}\lambda^{(i)}})
\ee
where $\alpha$ is a constant, the ``length'' of the integration contour $\gamma$.
We notice that this relation is nothing but the Weyl determinant formula relating characters to the Schur polynomials
\be
P_R(e^{\frac{\alpha}{\hbar}\lambda^{(i)}})=\chi_R\left(t_k=\frac{1}{k}\sum\lm_i e^{\frac{k\alpha}{\hbar}\lambda^{(i)}}\right)\\
\chi_R(t)=\det_{ij}s_{\lambda_i -i+j}(t),\quad e^{\sum\lm_k t_k z^k}=\sum\lm_k s_k(t)z^k
\ee
The integrals over $\gamma\in\Sigma_0$ of the eigenvalues $\lambda^{(i)}$ are thought of as integrals over different pre-image contours $\tilde \gamma$ on the spectral cover $\Sigma$ of a SW differential $\lambda$
\be
 \oint\lm_{\gamma}\lambda^{(i)}:=\oint\lm_{\tilde \gamma^{(i)}}\lambda,\quad \pi(\tilde \gamma^{(i)})=\gamma
\ee
And the exponent of the latter expression we define as a Fock-Goncharov coordinate (cluster variable):
\be
\X_{\tilde\gamma}:=e^{\frac{1}{\hbar}\oint\lm_{\tilde{\gamma}}\lambda}
\ee
Thus by analogy one writes an naive asymptotic form as
\be
\Tr_R \Pexp \frac{1}{\hbar}\oint\lm_{\gamma} A\sim \chi_R\left[t_k=\frac{1}{k}\sum\lm_{\tilde{\gamma}|\pi(\tilde{\gamma})=\gamma} w_{\tilde{\gamma}}^k\right]
\ee
The basic example is the line in the fundamental representation for $SU(2)$. In this case the cover is 2-fold, and the contour $\gamma$ has two pre-images $\tilde{\gamma}_1=\tilde{\gamma}$ and $\tilde{\gamma}_2=-\tilde{\gamma}$. The fundamental character reads then:
\be
\Tr_{[1]}\Pexp\frac{1}{\hbar}\oint\lm_{\gamma}A=\chi_{[1]}(t)=t_1=\X_{\tilde\gamma}+\X_{-\tilde\gamma}=\X_{\tilde\gamma}+\frac{1}{\X_{\tilde{\gamma}}}
\ee
This expression should be compared to the usual $sl_2$ character
\be
\Tr_{[1]}y^{\sigma_3}=y^{\frac{1}{2}}+\frac{1}{y^{\frac{1}{2}}}
\ee

{\it Notice} that this asymptotic form holds when one considers the quantized connection and the differential on the spectral curve, though it misses two important effects: the Stokes phenomenon and measure contributions described in s.\ref{sec:WKB} and s.\ref{sec:quantizingSC} respectively.

A similar approach taking into account instanton corrections from the quantum mechanics description of integrable systems is developed in \cite{Krefl}. Unfortunately, calculations are made in the Nekrasov-Shatashvili limit ($\hbar' = 0$ in our language), though non-perturbative Stokes' corrections are applied to construct an extra non-perturbative contribution to the prepotential in this limit. It is natural to assume consequent non-perturbative corrections to the Nekrasov partition function for ${\cal N}=2$ SUSY gauge theory.

A similar deformation of characters can be encountered under similar circumstances in \cite{NPS} (so called qq-characters).

To consider not only symmetric representations one needs to introduce multiple covers representing higher rank groups or multi-matrix models.

In such models eigenvalues $z_i^{(j)}$ acquire an extra ``flavour'' index $(j)$ running from 1 to $n-1$ for $SU(n)$. Thus one expects the following kind of expression for the determinant operator
\be
\check{\cal D}_{[r_1,\ldots,r_{n-1}]}(\zeta)\sim \left<\prod\lm_{j=1}^{n-1}\prod_{i}(\zeta-z_{i}^{(j)})^{r_j}\right>
\ee

\section{Knot invariants from WKB morphisms}

\subsection{Reidemeister invariants from quantum field theory}\label{sec:Reidemeister}

Let us consider the Chern-Simons theory with gauge group $G$ \cite{CS}, and the Wilson averages in this theory, which is knot polynomial can be associated with conformal blocks of two-dimensional conformal theory with positions of points changing in time \cite{Wit}. Since the theory is topological, one can consider just monodromies of the conformal blocks. The conformal theory that corresponds to this gauge theory is the Wess-Zumino-Witten-Novikov (WZWN) model \cite{WZWN} so that its correlators satisfy the Knizhnik-Zamolodchikov equation \cite{KZ}, and they can be considered as a wave function in Chern-Simons theory \cite{Moore-Seiberg}.

These picture of the Wilson averages allows one to connect knots with the Knizhnik-Zamolodchikov equation. Indeed, consider a knot on a 3-manifold $M_3={\cal C}\times [0,1]$ in a braid representation. The braid is given by $n$ trajectories $\gamma_i:\; z_i(t)\in{\cal C},\; t\in[0,1]$. The wave function on a time slice $t$ depends on the positions of the strands in the braid $z_i$. Consider a $\bigotimes_j Q_j\bigotimes_j \bar Q_j$ bundle over the configuration space ${\cal C}^n$ with connection
\be
{\cal D}_j =\p_{z_j}-A(z_j)
\ee

If the connection is flat
\be
\left[{\cal D}_i,{\cal D}_j\right]=0
\ee
one can construct the wave function as its flat section
\be
{\cal D}_i\Psi=0
\ee
The evolution operators can be interpreted as open Wilson lines in the ambient 3d theory:
\be
I=\Pexp \int\lm_0^1 dt \bigoplus_j A\left(z_j(t)\right)\dot z_j(t)
\ee
They are braid invariants (due to the flatness condition):
\be
\frac{\delta}{\delta \gamma_j(t)}I=0
\ee
The Wilson operators have a natural structure of the Hopf algebra. Correspondingly the space of wave functions can be endowed with the structure of a tensor category:
\be\label{M}
\bigotimes_j R_j=\bigoplus_{Q\vdash \sum\lm_j |R_j|}{\cal M}_Q\otimes Q
\ee
The Wilson operators diagonalize under this decomposition:
\be
I(\bigotimes_j R_j)=\bigoplus_{Q\vdash \sum\lm_j |R_j|}I({\cal M}_Q)\otimes \mathds{1}(Q)
\ee
There are two generating elements of the braid group which present two following cobordisms:
\begin{center}
\begin{tabular}{c|c|c|c}
  Cobordism&Trajectory & Representations & Diagram \\
  \hline
   $T_{i,i+1}$ &\includegraphics[scale=0.06]{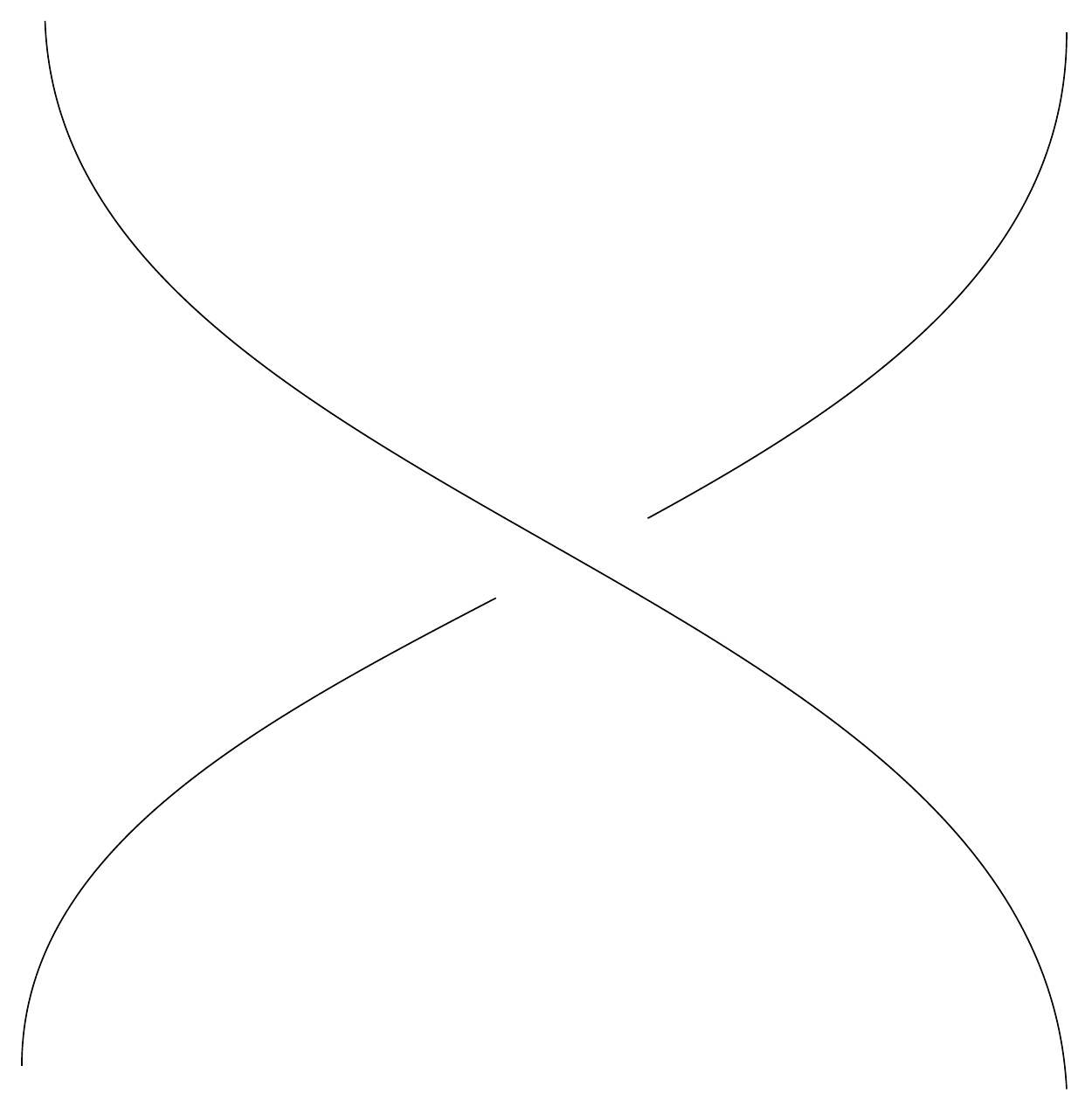}& $(V_1(z_1)\otimes V_2(z_2))\longrightarrow (V_1(z_2)\otimes V_2(z_1))$ & \includegraphics[scale=0.35]{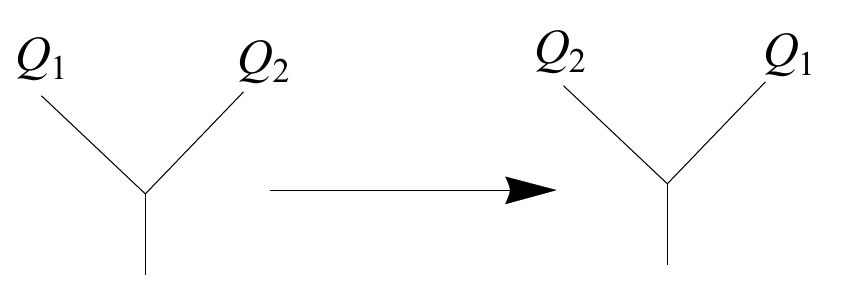} \\
\hline
   $S_{i,i+1,i+2}$ &
\includegraphics[scale=0.06]{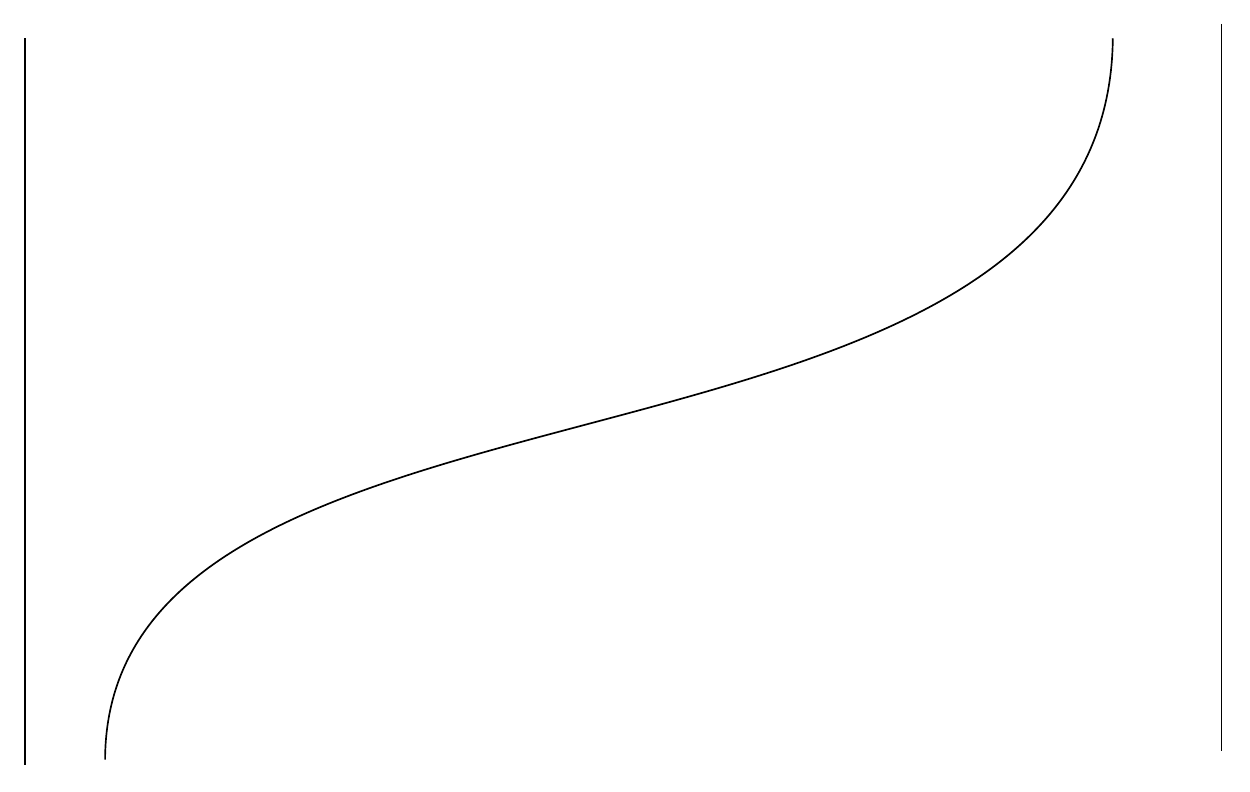}
 & $(V_1(z_1)\otimes V_2(z_2))\otimes V_3(z_3)\longrightarrow V_1(z_1)\otimes (V_2(z_2)\otimes V_3(z_3))$  &
\includegraphics[scale=0.35]{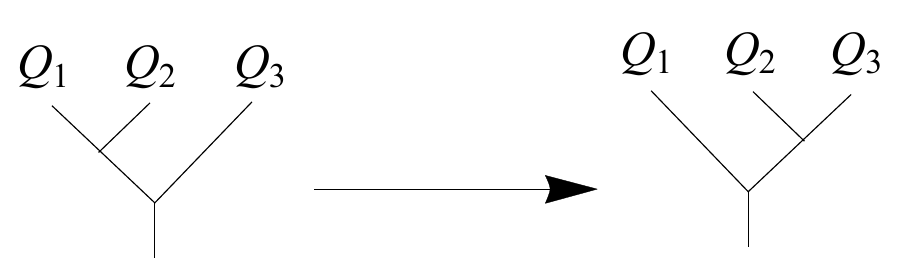} \\
\end{tabular}
\end{center}

\bigskip

Now in order to construct the knot invariant colored by a representation $Q$ from the braid, one has to construct a projector onto $Q$:
\be
{\cal P}_Q\left(\bigoplus_R {\cal M}_R\otimes R\right) ={\cal M}_Q\otimes Q
\ee
and then to ``remove'' the free ends of the braid either by taking trace or by gluing ``caps'' to its pairwise ends:
the wave functions with vacuum quantum numbers $\Psi_Q(z_1,z_2)\rightsquigarrow {\cal P}_\varnothing (Q\otimes Q)$.

\bigskip

There is another set of important operators which inserting the Wilson line into the fixed time slice. Since the lines are completely within the slice, these operators are colorless and act on the wave functions as
\be
W_Q=\Tr_Q \left[\Pexp\oint A(z)dz\; Q\otimes\right]={\rm Unknot}_Q\; {\cal P}_\varnothing \left(Q\otimes \Pexp\oint A(z)dz\; Q\right) \otimes
\ee
By trace here we mean only the trace over the representation of the gauge group, hence, the operator is a well-defined scalar.
\begin{figure}[htbp]
\begin{center}
\includegraphics[scale=0.5]{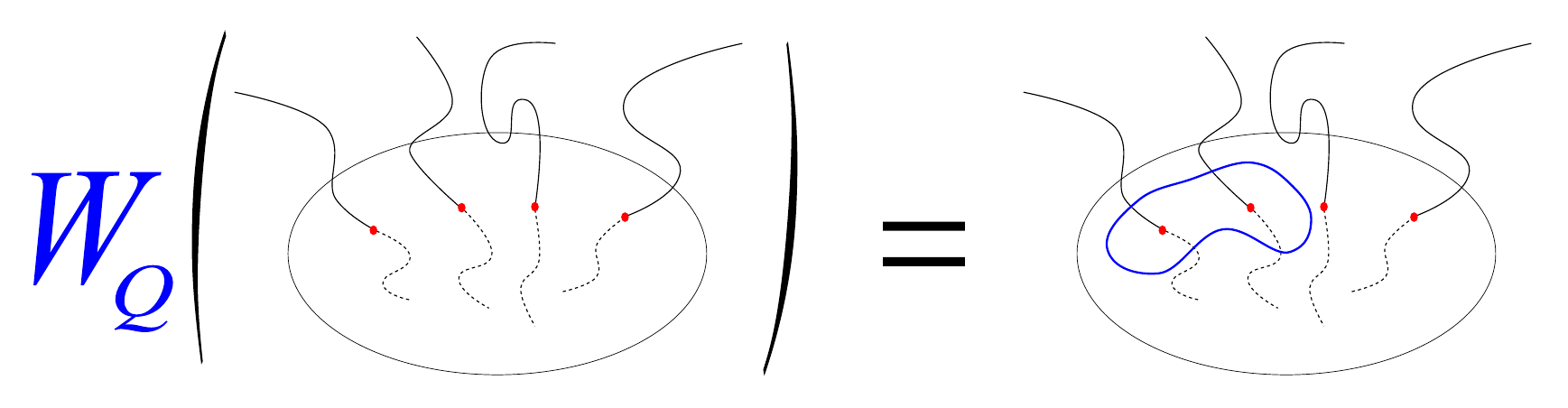}
\caption{$W_Q$\label{fig:W_Q}}
\end{center}
\end{figure}
Decorating knots with operations of this kind was considered in detail in \cite{Morton}.
In fact, one can easily understand that for the one-point conformal block (or for the operator $W_Q$ that inserts a loop surrounding only one point)
\be
W_Q \Psi_{Q'}={H_{Q,Q'}\over\hbox{dim}_q(Q')} \Psi_{Q'}
\ee
where $H_{Q,Q'}$ is the Hopf link HOMFLY polynomial and dim$_q(Q)$ is the quantum dimension of $Q$, which is the same as the (unreduced=non-normalized) HOMFLY polynomial of the unknot.
If now one applies such an operator to a few strands, it can be expanded into the cluster coordinates which are a kind of basis in the Hilbert space. Then, generalizing \cite{surop,GMM} one has to expect that $W_Q$ is a polynomial in these coordinates which is a character in full analogy with s.4.4,
\be
W_Q=\chi_Q (Y_1,\ldots,Y_k)
\ee
and $Y_k$ are the Darboux coordinates on the $\cal C$ flat connections moduli space.

In the next two subsections we present two different realizations of the described scheme, which is basically related with two different realizations of ${\cal R}$-matrices.

\subsection{Knot invariants from RTW representation via duality kernels}

\subsubsection{The basic idea}

One of the possibilities to realize this general construction due to E.Witten \cite{Wit} is the Reshetikhin-Turaev scheme \cite{RT}, which was realized in detail in \cite{inds} and in \cite{MMMI,MMMII} for different braid representations of knots/links. The approach is based on assigning with any cross of the braid the ${\cal R}$-matrix of $U_q(G)$. This ${\cal R}$-matrix can either come as a monodromy (modular) matrix of the WZWN theory \cite{inds} or can be treated as just a numeric ${\cal R}$-matrix from representation theory \cite{MMMII}. Here we will propose a third possibility: to reproduce the ${\cal R}$-matrix by the modular kernel of the conformal field theory. Since this case is described by the Virasoro algebra, the obtained ${\cal R}$-matrix is associated with $SU_q(2)$ and the corresponding knot invariants are the Jones polynomials.

 Let us explain how to apply the modular transformations to evaluation of the knot polynomials. The idea is that, if there are three strands, one can describe the crossing of the first two strands and the second and the third strands correspondingly as
\be
\boxed{
{\cal R}\otimes I = T , \ \ \ \ \ \ \   I\otimes {\cal R}= STS^\dagger
}
\ee
i.e. the modular matrix $T$ plays the role of the ${\cal R}$-matrix acting in the space of intertwining operators ${\cal M}_Q$ (\ref{M}) and $S$ plays the role of the mixing matrix in the RTW formalism, see details in \cite{MMMII}. These transformations, $S$ and $T$, are known to form a Moore-Seiberg grouppoid \cite{Moore-Seiberg,PT}.

Now one has to check the Reidemeister moves:

\begin{itemize}
\item 3-rd Reidemeister = YB relation
\be
\Big(I\otimes R\Big)\Big(R\otimes I\Big)\Big(I\otimes R\Big) =
\Big(R\otimes I\Big)\Big(I\otimes R\Big)\Big(R\otimes I\Big)
\ee
i.e.
\be
 STS^\dagger T STS^\dagger = TSTS^\dagger T
\ee
is solved by the anzatz
\be
SS^\dagger =1, \nn \\
STS^\dagger TST =I
\label{STrels0}
\ee
because it can be rewritten as
\be
\Big(STS^\dagger T ST\Big)S^\dagger = T\Big(STS^\dagger TST\Big)T^{-1}S^{-1}
\ee

In the simplest situation (the four-point spherical conformal block)
we additionally have $S^\dagger = S$, and therefore (\ref{STrels0})
reduce to
\be
S^2=1 \ \ \ \ {\rm  and} \ \ \ \  (ST)^3=1
\label{STcube}
\ee

\item 2-nd Reidemeister:  $TT^{-1}=1$

\item 1-st Reidemeister:
\be
T^{ij}_{kl} P^k_i = P^j_l
\ee
where $P$ is the cap projector.
\end{itemize}

\bigskip

The simplest non-trivial solution to (\ref{STcube}) is in $2\times 2$ matrices.
If $T$ is diagonalized, then $S$ is the elementary mixing matrix of \cite{MMMII}:
\be
T = \left(\begin{array}{cc}  q & 0 \\ \\ 0 & -\frac{1}{q}\end{array}\right), \ \ \ \ \ \
S = \left(\begin{array}{cc}  \frac{1}{[2]} & \frac{\sqrt{[3]}}{[2]} \\ \\
\frac{\sqrt{[3]}}{[2]} & -\frac{1}{[2]}\end{array}\right)
\label{ST2}
\ee
where quantum numbers $[2]=q+q^{-1}$ and $[3] = q^2+1+q^{-2}$.

\subsubsection{ $S$ in (\ref{ST2}) as the Racah matrix}

Consider the representation-product diagrams from \cite{MMMII} for the particular
choice of external legs:

\begin{picture}(300,100)(-150,-30)
\put(0,0){\line(0,-1){20}}
\put(0,0){\line(-1,1){40}}
\put(0,0){\line(1,1){40}}
\put(-20,20){\line(1,1){20}}
\put(200,0){\line(0,-1){20}}
\put(200,0){\line(-1,1){40}}
\put(200,0){\line(1,1){40}}
\put(220,20){\line(-1,1){20}}
\put(5,-22){\mbox{$[1]$}}
\put(205,-22){\mbox{$[1]$}}
\put(-46,46){\mbox{$[1]$}}
\put(-4,46){\mbox{$[1]$}}
\put(38,46){\mbox{$\overline{[1]}$}}
\put(154,46){\mbox{$[1]$}}
\put(196,46){\mbox{$[1]$}}
\put(238,46){\mbox{$\overline{[1]}$}}
\put(-15,2){\mbox{$i$}}
\put(212,2){\mbox{$J$}}
\put(40,13){\mbox{$= \ \ \ \sum_J \  S_{iJ}\left[
\begin{array}{cc} \l[1] & \l[1] \\ &\\\overline{[1]} & \overline{[1]}\end{array}
\right]$}}
\end{picture}

\noindent
Both sets of intermediate states are $2$-dimensional, but different:
$i=[2],[11]$ and $J=0,Adj$, i.e. $J=[0],[21^{N-1}]$. Note also that the
conjugate fundamental representation $\overline{[1]} = [1^{N-1}]$.
For $N=2$ there are coincidences: $\overline{[1]}=[1]$, $Adj=[2]$, $[11]=[0]$,
therefore the two diagrams are the same, moreover the matrix $S_{iJ}$
coincides with that for the Racah matrix for $[1]^{\otimes 3}\longrightarrow [1]$,
which is known from ref.\cite{MMMII} to be exactly (\ref{ST2}).

\subsubsection{$S$ and $T$ matrices from conformal theory}

Instead of trying to find solutions to eqs.(\ref{STcube}) within the group theory framework, one can use another possibility: the same equations are solved by the modular kernels that control modular transformations of the conformal blocks. In the generic case, these transformations are given by integral kernels. However, in the case degenerate fields they become matrices. Since the Virasoro algebra is associated with $SU(2)$, one expects obtained in this way the $S$- and $T$-matrices to generate the colored Jones polynomials, while going further to $SU(N)$ with $N>2$ would require modular transformations of the conformal blocks of the corresponding $W_N$-algebras.

Thus, we are going to consider the conformal blocks with the fields $\Phi_{(m,n)}$ degenerate at level $m\cdot n$ which have conformal dimensions \cite{BPZ}
\be\label{Kac}
\Delta_{(m,n)}=\alpha_{(m,n)} \Big(\alpha_{(m,n)} -b+{1\over b}\Big),\ \ \ \ \ \ \ \
 \alpha_{(m,n)}={1\over 2}\Big({m-1\over b}-(n-1)b\Big)
\ee
and $b$ parameterizes the central charge of conformal theory: $c=1-6(b-1/b)^2$.
At the same time, choosing different $n$ changes the spin of representation (of the colored Jones polynomial).

Now we are going to read off the matrix $S$ from the modular transformation
\be
B_{j_s}\left[\begin{array}{cc}
    j_2 & j_3\\
    j_1 & j_4\\
\end{array}\right](x)=\sum\limits_{j_t} S_{j_s\; j_t}\left[\begin{array}{cc}
j_2 & j_3\\
j_1 & j_4\\
\end{array}\right]B_{j_t}\left[\begin{array}{cc}
j_2 & j_1\\
j_3 & j_4\\
\end{array}\right](1-x)
\ee

\paragraph{The fundamental representation.} Let us consider the simplest example of the fundamental representation of $SU(2)$. In this case, one may expect that the end of the Wilson line in the fundamental representation $[1]$ behaves as $\Phi_{(1,2)}$ in the conformal theory
\be
\Phi_{(1,2)}\otimes\Phi_{(1,2)}=\Phi_{(1,3)}\oplus\Phi_{(1,1)}
\ee
Then, one has (where index means a projection in the intermediate channel on the corresponding state)
\be
B_{[11]}(x)=\left.\left\langle \Phi_{(1,2)}(0)\Phi_{(1,2)}(x)\right|_{(1,1)}\Phi_{(1,2)}(1)\Phi_{(1,2)}(\infty)\right\rangle =x^{\delta}(1-x)^{\delta}{}_2F_1\left[\begin{array}{c}
	\alpha,\; \beta \\
	\gamma
\end{array}
\right](x),\\
B_{[2]}(x)=\left.\left\langle \Phi_{(1,2)}(0)\Phi_{(1,2)}(x)\right|_{(1,3)}\Phi_{(1,2)}(1)\Phi_{(1,2)}(\infty)\right\rangle=x^{\bar\delta}(1-x)^{\delta}{}_2F_1\left[\begin{array}{c}
	\alpha-\gamma+1,\; \beta-\gamma+1 \\
	2-\gamma
\end{array}
\right](x)
\ee
where
\be
\alpha=\frac{3}{2 b^2}+\frac{7}{2},\quad \beta=\frac{1}{2 b^2}+\frac{3}{2},\quad \gamma=\frac{1}{b^2}+3,\quad \delta=\frac{3 b^2}{4}+\frac{1}{4 b^2}+\frac{3}{2},\quad\bar\delta=\frac{3 b^2}{4}-\frac{3}{4 b^2}-\frac{1}{2}
\ee
Then, the matrix $S$ reads in this case ($S^2= 1$)
\be
S=\left(
\begin{array}{cc}
	\frac{\Gamma \left(2+\frac{2}{b^2}\right) \Gamma \left(-\frac{2}{b^2}-1\right)}{\Gamma \left(1+\frac{1}{b^2}\right) \Gamma \left(-\frac{1}{b^2}\right)} & \frac{\Gamma \left(2+\frac{2}{b^2}\right) \Gamma \left(\frac{2}{b^2}+1\right)}{\Gamma \left(1+\frac{1}{b^2}\right) \Gamma \left(\frac{3}{b^2}+2\right)} \\
	\frac{\Gamma \left(-\frac{2}{b^2}\right) \Gamma \left(-\frac{2}{b^2}-1\right)}{\Gamma \left(-\frac{1}{b^2}\right) \Gamma \left(-\frac{3}{b^2}-1\right)} & \frac{\Gamma \left(-\frac{2}{b^2}\right) \Gamma \left(\frac{2}{b^2}+1\right)}{\Gamma \left(-\frac{1}{b^2}\right) \Gamma \left(\frac{1}{b^2}+1\right)} \\
\end{array}
\right)=e^{i\pi}\left(
\begin{array}{cc}
	{1\over [2]} & {\sqrt{[3]}\over [2]}\gamma^2 \\
{\sqrt{[3]}\over [2]}\gamma^{-2} 	& -{1\over [2]} \\
\end{array}
\right)=U\cdot \left(
\begin{array}{cc}
	{1\over [2]} & {\sqrt{[3]}\over [2]} \\
{\sqrt{[3]}\over [2]} 	& -{1\over [2]} \\
\end{array}
\right)\cdot U^{-1}
\ee
where
\be
\gamma^2=(2+b^2){[2]\over\sqrt{[3]}}{\Gamma^2\left(1+{2\over b^2}\right)\over \Gamma\Big({1\over b^2}\Big)\Gamma\Big(2+{3\over b^2}\Big)}
\ee
and
\be\label{U}
U=e^{{i\pi\over 2}}\left(
\begin{array}{cc}
	\gamma & 0 \\
0 	& \gamma^{-1} \\
\end{array}
\right)
\ee
while the matrix $T$ is\footnote{In this section $q=e^{\pi i b^{-2}}$.} ($(ST)^3\sim 1$)
\be
T\sim
\left(
\begin{array}{cc}
	q & 0 \\
	0 & -{1\over q} \\
\end{array}
\right)
\ee
The overall normalization of the matrix $T$ is an inessential overall state space phase, it can be fixed from the requirement $(S T)^3=1$. When checking various relations here, we used
\be
\Gamma(z)\Gamma(1-z)=\frac{\pi}{\sin \pi z}, \quad \cos(\pi \alpha b^{-2})=\frac{q^{\alpha}+q^{-\alpha}}{2}, \quad \sin(\pi \alpha b^{-2})=\frac{q^{\alpha}-q^{-\alpha}}{2i}
\ee

Note that this matrix $S$ differs from that in (\ref{ST2}) by the additional $U$-conjugation (\ref{U}). However, this conjugation influences neither on the relation $(ST)^3=1$, nor on $S^2=1$, and does not change the answers for the knot polynomials.

\paragraph{Higher spin representations.}

So far we considered only the fundamental representation of $SU(2)$. Similarly, one can consider representations of higher spins. To this end, one has to use $\Phi_{(1,2)}\otimes \Phi_{(1,2)}\otimes \Phi_{(1,s+1)}\otimes \Phi_{(1,s+1)}$ with the fusion matrices
\be
S=\left(
\begin{array}{cc}
	\frac{\Gamma \left(2+\frac{2}{b^2}\right) \Gamma \left(-\frac{s+2}{b^2}+\frac{1}{b^2}-1\right)}{\Gamma \left(1+\frac{1}{b^2}\right) \Gamma \left(-\frac{s+2}{b^2}+\frac{2}{b^2}\right)} & \frac{\Gamma \left(2+\frac{2}{b^2}\right) \Gamma \left(\frac{s+2}{b^2}-\frac{1}{b^2}+1\right)}{\Gamma \left(1+\frac{1}{b^2}\right) \Gamma \left(\frac{s+2}{b^2}+2\right)} \\
	\frac{\Gamma \left(-\frac{2}{b^2}\right) \Gamma \left(-\frac{s+2}{b^2}+\frac{1}{b^2}-1\right)}{\Gamma \left(-\frac{1}{b^2}\right) \Gamma \left(-\frac{s+2}{b^2}-1\right)} & \frac{\Gamma \left(-\frac{2}{b^2}\right) \Gamma \left(\frac{s+2}{b^2}-\frac{1}{b^2}+1\right)}{\Gamma \left(-\frac{1}{b^2}\right) \Gamma \left(\frac{s+2}{b^2}-\frac{2}{b^2}+1\right)} \\
\end{array}
\right)\\
T=\left(\begin{array}{cc}
	q^{\frac{s+1}{2}}e^{\frac{i\pi}{2}} & 0\\
	0 & q^{-\frac{s+1}{2}}e^{-\frac{i\pi}{2}}\\
\end{array}\right)
\ee
These matrices can be obtained either directly from the equations for the degenerate conformal fields, or from the general expression for the modular kernel due to B.Ponsot and J.Teschner \cite{PT}. This latter procedure is discussed in the Appendix.

Using these $S$ and $T$ matrices, one can easily generate the Jones polynomials as it was explained above. Note that one can easily construct the most generic modular kernel $S$, when the only field is degenerate at the second level:
the general degenerate conformal block with $j_2=1$ is described by the hypergeometric function
\be
B(x)\sim{}_2F_1 \left[\begin{array}{c}
2+\frac{b^{-2}}{2}(3+j_1+j_3+j_4)\quad 1+\frac{b^{-2}}{2}(1+j_1+j_3-j_4)\\
2+b^{-2}(1+j_1)
\end{array}\right](x)
\ee
The corresponding monodromy matrix reads\footnote{To be precise, this matrix is related to the modular kernel as
\be
 S_{j_s\; j_t}\left[\begin{array}{cc}
    1 & j_3\\
    j_1 & j_4\\
 \end{array}\right]=\sum\limits_{h,h'=\pm 1}\delta(j_s-j_1-h)\delta(j_t-j_3-h')S(j_1,1,j_3,j_4)_{h,h'}
\ee}
\be
S(j_1,1,j_3,j_4)=\left(
\begin{array}{cc}
    \frac{\Gamma \left(\frac{{j_1}+1}{b^2}+2\right) \Gamma \left(-\frac{b^2+{j_3}+1}{b^2}\right)}{\Gamma \left(\frac{2 b^2+ {j_1}- {j_3}+ {j_4}+1}{2 b^2}\right) \Gamma \left(-\frac{1- {j_1}+ {j_3}+ {j_4}}{2 b^2}\right)} & \frac{\Gamma \left(\frac{ {j_1}+1}{b^2}+2\right) \Gamma \left(\frac{b^2+ {j_3}+1}{b^2}\right)}{\Gamma \left(\frac{2 b^2+ {j_1}+ {j_3}- {j_4}+1}{2 b^2}\right) \Gamma \left(\frac{4 b^2+ {j_1}+ {j_3}+ {j_4}+3}{2 b^2}\right)} \\
    \frac{\Gamma \left(-\frac{ {j_1}+1}{b^2}\right) \Gamma \left(-\frac{b^2+ {j_3}+1}{b^2}\right)}{\Gamma \left(-\frac{ {j_1}+ {j_3}- {j_4}+1}{2 b^2}\right) \Gamma \left(-\frac{2 b^2+ {j_1}+ {j_3}+ {j_4}+3}{2 b^2}\right)} & \frac{\Gamma \left(-\frac{ {j_1}+1}{b^2}\right) \Gamma \left(\frac{b^2+ {j_3}+1}{b^2}\right)}{\Gamma \left(-\frac{ {j_1}- {j_3}+ {j_4}+1}{2 b^2}\right) \Gamma \left(\frac{2 b^2- {j_1}+ {j_3}+ {j_4}+1}{2 b^2}\right)} \\
\end{array}
\right)
\ee
The higher spin matrix is generated from the recursion formula derived \cite{Go} from the ``cabling'' procedure ($[r]\otimes [1]=[r+1]\oplus [r-1]$):
\be
S_{q,q'}\left[\begin{array}{cc}
	r+1 & j_3 \\
	j_1 & j_4 		
\end{array}\right]=\sum\lm_{s,p}S_{r+1,s}\left[\begin{array}{cc}
1 & q \\
r & j_1 		
\end{array}\right]
S_{q,p}\left[\begin{array}{cc}
	1 & j_3 \\
	s & j_4 		
\end{array}\right]
S_{s,q'}\left[\begin{array}{cc}
	r & p \\
	j_1 & j_4 		
\end{array}\right]
S_{p,r+1}\left[\begin{array}{cc}
	r & 1 \\
	q' & j_3 		
\end{array}\right]
\ee

\subsubsection{Plat representation of link diagrams (spherical conformal block)}

Since operators $S$ and $T$, satisfying (\ref{STrels0}) naturally arise
as modular transformations of conformal blocks,
one can associate them with the link diagrams in the plat representation
in the following way.

We begin with examples.

\paragraph{ 1 cap}
There is nothing to consider in the case of one cap:
independently of the number of interweavings between two strands,
it is always the unknot:
\be
(T^{n})_{22} = (-q)^{-n}
\ee

\paragraph{ 2 caps}

Our notation should be clear from the picture, where the bottom pictures present the conformal block one starts with, the middle pictures present the monodromy of points in the conformal block and the top pictures present the resulting conformal block:

\begin{picture}(100,190)(-50,-80)
\put(0,0){\vector(0,-1){20}}
\put(20,-20){\vector(0,1){20}}
\qbezier(0,-20)(0,-30)(10,-30)
\qbezier(20,-20)(20,-30)(10,-30)
\put(80,0){\vector(0,-1){20}}
\put(60,-20){\vector(0,1){20}}
\qbezier(60,-20)(60,-30)(70,-30)
\qbezier(80,-20)(80,-30)(70,-30)
\put(10,-50){\line(1,0){60}}
\put(10,-50){\line(-2,1){20}}
\put(10,-50){\line(-2,-1){20}}
\put(70,-50){\line(2,1){20}}
\put(70,-50){\line(2,-1){20}}
\put(38,-47){\mbox{$0$}}
\qbezier(0,0)(0,15)(10,20)
\qbezier(10,20)(20,25)(20,40)
\qbezier(20,0)(20,15)(10,20)
\qbezier(10,20)(0,25)(0,40)
\put(10,60){\line(1,0){60}}
\put(10,60){\line(-2,1){20}}
\put(10,60){\line(-2,-1){20}}
\put(70,60){\line(2,1){20}}
\put(70,60){\line(2,-1){20}}
\put(38,63){\mbox{$0$}}
\qbezier(-18,68)(-22,60)(-18,52)
\put(-18.5,67){\vector(1,2){2}}
\put(-18.5,53){\vector(1,-2){2}}
\put(200,0){\vector(0,-1){20}}
\put(220,-20){\vector(0,1){20}}
\qbezier(200,-20)(200,-30)(210,-30)
\qbezier(220,-20)(220,-30)(210,-30)
\put(300,0){\vector(0,-1){20}}
\put(280,-20){\vector(0,1){20}}
\qbezier(280,-20)(280,-30)(290,-30)
\qbezier(300,-20)(300,-30)(290,-30)
\put(210,-50){\line(1,0){80}}
\put(210,-50){\line(-2,1){20}}
\put(210,-50){\line(-2,-1){20}}
\put(290,-50){\line(2,1){20}}
\put(290,-50){\line(2,-1){20}}
\put(248,-47){\mbox{$0$}}
\qbezier(220,0)(220,15)(230,20)
\qbezier(230,20)(240,25)(240,40)
\qbezier(280,0)(280,15)(270,20)
\qbezier(270,20)(260,25)(260,40)
\put(250,60){\line(0,1){20}}
\put(250,60){\line(-2,-1){20}}
\put(250,60){\line(2,-1){20}}
\put(250,80){\line(-2,1){20}}
\put(250,80){\line(2,1){20}}
\put(254,67){\mbox{$j$}}
%
%
\put(-12,-25){\mbox{$\bar r_1$}}
\put(23,-25){\mbox{$r_1$}}
\put(48,-25){\mbox{$r_2$}}
\put(83,-25){\mbox{$\bar r_2$}}
\put(-22,-65){\mbox{$\bar r_1$}}
\put(-22,-42){\mbox{$r_1$}}
\put(95,-42){\mbox{$r_2$}}
\put(95,-65){\mbox{$\bar r_2$}}
\put(188,-25){\mbox{$\bar r_1$}}
\put(223,-25){\mbox{$r_1$}}
\put(268,-25){\mbox{$ r_2$}}
\put(303,-25){\mbox{$\bar r_2$}}
\put(178,-65){\mbox{$\bar r_1$}}
\put(178,-42){\mbox{$r_1$}}
\put(315,-42){\mbox{$r_2$}}
\put(315,-65){\mbox{$\bar r_2$}}
\put(215,50){\mbox{$\bar r_1$}}
\put(215,90){\mbox{$r_1$}}
\put(275,90){\mbox{$r_2$}}
\put(275,50){\mbox{$\bar r_2$}}
\put(200,70){\mbox{$\bigoplus_j$}}
\end{picture}

Expressions for the two operations are respectively $T_0(\bar r_1, r_1)$
and $S_{j0}\left(\begin{array}{cc} r_1 & \bar r_2 \\ \bar r_1 & r_2\end{array}\right)$.

Generic knot/link in this sector is a sequence of $T$-twists between parallel strands
in the channel $23$ and antiparallel strands in the channel $12$ (numbers $1,2,3,4$ here
label the vertical lines in the picture).
This family includes 2-strand links and knots, twist knots, antiparallel 2-strand links,
double braids from \cite{evo} and in general is known as a family of 2-bridge links.

\begin{itemize}

\item {\bf Two unknots}

If no $S$ operators are applied, we get two disconnected unknots.
The answer for two fundamental representations of $SU(2)$ is obtained with the help of
(\ref{ST2}):
\be
\big(T^{n_1}\big)_{22} \big(T^{n_2}\big)_{22} = (-q)^{-n_1-n_2}
\ee
Up to the framing factor this is the fully reduced knot polynomial
(i.e. unreduced expression $[2]^2$ is divided by square of the quantum dimension $[2]$).

\bigskip

\item {\bf 2-strand torus links}

The plat diagram and the sequence of modular transformations in this case are:

\begin{picture}(300,280)(-50,-50)
\put(0,0){\vector(0,-1){20}}
\put(20,-20){\vector(0,1){20}}
\qbezier(0,-20)(0,-30)(10,-30)
\qbezier(20,-20)(20,-30)(10,-30)
\put(100,0){\vector(0,-1){20}}
\put(80,-20){\vector(0,1){20}}
\qbezier(80,-20)(80,-30)(90,-30)
\qbezier(100,-20)(100,-30)(90,-30)
\qbezier(20,0)(20,15)(30,20)
\qbezier(30,20)(40,25)(40,40)
\qbezier(80,0)(80,15)(70,20)
\qbezier(70,20)(60,25)(60,40)
\qbezier(40,40)(40,55)(50,60)
\qbezier(50,60)(60,65)(60,80)
\qbezier(60,40)(60,55)(50,60)
\qbezier(50,60)(40,65)(40,80)
\put(43,90){\mbox{$\ldots$}}
\qbezier(40,100)(40,115)(50,120)
\qbezier(50,120)(60,125)(60,140)
\qbezier(60,100)(60,115)(50,120)
\qbezier(50,120)(40,125)(40,140)
\qbezier(20,180)(20,165)(30,160)
\qbezier(30,160)(40,155)(40,140)
\qbezier(80,180)(80,165)(70,160)
\qbezier(70,160)(60,155)(60,140)
\put(0,200){\vector(0,-1){220}}
\put(20,180){\vector(0,1){20}}
\qbezier(0,200)(0,210)(10,210)
\qbezier(20,200)(20,210)(10,210)
\put(100,200){\vector(0,-1){220}}
\put(80,180){\vector(0,1){20}}
\qbezier(80,200)(80,210)(90,210)
\qbezier(100,200)(100,210)(90,210)
\put(210,-15){\line(1,0){80}}
\put(210,-15){\line(-2,1){20}}
\put(210,-15){\line(-2,-1){20}}
\put(290,-15){\line(2,1){20}}
\put(290,-15){\line(2,-1){20}}
\put(248,-10){\mbox{$0$}}
\put(210,195){\line(1,0){80}}
\put(210,195){\line(-2,1){20}}
\put(210,195){\line(-2,-1){20}}
\put(290,195){\line(2,1){20}}
\put(290,195){\line(2,-1){20}}
\put(248,200){\mbox{$0$}}

\put(250,60){\line(0,1){20}}
\put(250,60){\line(-2,-1){20}}
\put(250,60){\line(2,-1){20}}
\put(250,80){\line(-2,1){20}}
\put(250,80){\line(2,1){20}}
\put(254,67){\mbox{$j$}}
\qbezier(234,95)(250,110)(266,95)
\put(236,97){\vector(-1,-1){2}}
\put(264,97){\vector(1,-1){2}}
\put(246,110){\mbox{$T_j^{2k}$}}
\put(-12,-25){\mbox{$\bar r_1$}}
\put(23,-25){\mbox{$r_1$}}
\put(68,-25){\mbox{$r_2$}}
\put(103,-25){\mbox{$\bar r_2$}}
%
%
%
\put(178,-30){\mbox{$\bar r_1$}}
\put(178,-7){\mbox{$r_1$}}
\put(315,-7){\mbox{$r_2$}}
\put(315,-30){\mbox{$\bar r_2$}}
\put(215,50){\mbox{$\bar r_1$}}
\put(215,90){\mbox{$r_1$}}
\put(275,90){\mbox{$r_2$}}
\put(275,50){\mbox{$\bar r_2$}}
\put(200,70){\mbox{$\bigoplus_j$}}
\put(178,208){\mbox{$r_1$}}
\put(178,180){\mbox{$\bar r_1$}}
\put(315,180){\mbox{$\bar r_2$}}
\put(315,208){\mbox{$  r_2$}}
\put(250,10){\vector(0,1){30}}
\put(250,140){\vector(0,1){30}}
\put(257,20){\mbox{$S_{j0}$}}
\put(257,150){\mbox{$S_{0j}$}}
\end{picture}

The corresponding analytical expression is
\be
\sum_j S_{0j}\left[\begin{array}{cc}\bar r_1 & \bar r_2\\r_1&r_2\end{array}\right]
\ T_j\big[r_1,r_2\big]^{2k}\
S_{j0}\left[\begin{array}{cc} r_1 & r_2 \\ \bar r_1 & \bar r_2 \end{array}\right]
\ee
In the case of two fundamental representations, $r_1=r_2=[1]$ of $SU(2)$,
one can use matrices (\ref{ST2})
and obtain:
\be
ST^{2k}S \ \stackrel{(\ref{ST2})}{=}
\ \frac{1}{[2]^2}\left(\begin{array}{cc} q^{2k}+q^{-2k}[3] & \big(q^{2k}-q^{-2k}\big)\sqrt{[3]} \\ \\
\big(q^{2k}-q^{-2k}\big)\sqrt{[3]} & q^{2k}[3]+q^{-2k} \end{array}\right)
\label{2straN2}
\ee
Expression in the right lower corner (the matrix element $22$) is exactly the {\it reduced} Jones polynomial
\be
J_{[1],[1]}^{[2,2k]} = \left.\frac{1}{[N]^2}\left(q^{2k}\frac{[N][N+1]}{[2]}+q^{-2k}\frac{[N][N-1]}{[2]}\right)
\right|_{N=2} = \frac{1}{[2]^2}\Big([3]q^{2k}+q^{-2k}\Big)
\label{2strlinkJones}
\ee
for the 2-strand torus links (in the Rosso-Jones framing \cite{RS}).

\bigskip

\item {\bf 2-strand torus knots}

The only difference in this case is that even power $2k$ is substituted by the odd one $2k+1$,
but this is only possible for two coincident representations $r_1=r_2$.
This restriction is obvious from the plat diagram on the left side of the above picture,
on the right side, one would get as the top picture the diagram

\begin{picture}(100,50)(-150,-75)
\put(10,-50){\line(1,0){60}}
\put(10,-50){\line(-2,1){20}}
\put(10,-50){\line(-2,-1){20}}
\put(70,-50){\line(2,1){20}}
\put(70,-50){\line(2,-1){20}}
\put(38,-47){\mbox{$0$}}
\put(-22,-65){\mbox{$\bar r_1$}}
\put(-22,-42){\mbox{$r_2$}}
\put(95,-42){\mbox{$r_1$}}
\put(95,-65){\mbox{$\bar r_2$}}
\end{picture}

\noindent
and again this is possible (the singlet can run in the intermediate line)
only if $r_1=r_2$.

As to formula (\ref{2straN2}), it remains just the same, with the obvious change
$2k\longrightarrow 2k+1$, and the 22 element of the matrix
reproduces the reduced Jones polynomial
\be
ST^{2k+1}S \ \stackrel{(\ref{ST2})}{=}
\ \frac{1}{[2]^2}\left(\begin{array}{cc} q^{2k+1}-q^{-2k-1}[3] & (q^{2k+1}-q^{-2k-1})\sqrt{[3]} \\ \\
(q^{2k+1}-q^{-2k-1})\sqrt{[3]} & q^{2k+1}[3]-q^{-2k-1} \end{array}\right) =
\ \left(\begin{array}{ccc} \ldots && \ldots \\ \\ \ldots &&
\frac{1}{[2]}J_{[1]}^{[2,2k+1]} \end{array}\right)
\label{2straN2kn}
\ee
where
\be
J_{[1]}^{[2,2k+1]} = \left.\frac{1}{[N]}\left(q^{2k+1}\frac{[N][N+1]}{[2]}-q^{-2k-1}\frac{[N][N-1]}{[2]}\right)
\right|_{N=2} = \frac{1}{[2]}\Big([3]q^{2k+1}-q^{-2k-1}\Big)
\ee
Note that in variance with links only one of the two factors $[2]$
is eliminated by expressing the answer through the
{\it reduced} knot polynomial.
Also note that, like in (\ref{2strlinkJones}),
the Jones polynomial appeared in the Rosso-Jones framing rather than in the topological one.

\bigskip

\item {\bf Twist knots} differ by insertion of two additional twists in the  channel $12$:

\begin{picture}(300,380)(-50,-40)
\put(0,0){\vector(0,-1){20}}
\put(20,-20){\vector(0,1){20}}
\qbezier(0,-20)(0,-30)(10,-30)
\qbezier(20,-20)(20,-30)(10,-30)
\put(80,0){\vector(0,-1){20}}
\put(100,-20){\vector(0,1){20}}
\qbezier(80,-20)(80,-30)(90,-30)
\qbezier(100,-20)(100,-30)(90,-30)
\qbezier(20,0)(20,15)(30,20)
\qbezier(30,20)(40,25)(40,40)
\qbezier(80,0)(80,15)(70,20)
\qbezier(70,20)(60,25)(60,40)
\qbezier(40,40)(40,55)(50,60)
\qbezier(50,60)(60,65)(60,80)
\qbezier(60,40)(60,55)(50,60)
\qbezier(50,60)(40,65)(40,80)
\put(43,90){\mbox{$\ldots$}}
\qbezier(40,100)(40,115)(50,120)
\qbezier(50,120)(60,125)(60,140)
\qbezier(60,100)(60,115)(50,120)
\qbezier(50,120)(40,125)(40,140)
\qbezier(20,180)(20,165)(30,160)
\qbezier(30,160)(40,155)(40,140)
%
%
\put(0,310){\vector(0,-1){53}}
\put(20,260){\vector(0,1){2}}
\qbezier(40,300)(40,310)(50,310)
\qbezier(60,300)(60,310)(50,310)
\put(60,300){\vector(0,-1){160}}
%
\qbezier(0,310)(0,320)(10,320)
\qbezier(100,310)(100,320)(90,320)
\put(90,320){\vector(-1,0){80}}
\put(100,310){\line(0,-1){310}}
\qbezier(0,180)(0,195)(10,200)
\qbezier(10,200)(20,205)(20,220)
\qbezier(20,180)(20,195)(10,200)
\qbezier(10,200)(0,205)(0,220)
\qbezier(0,220)(0,235)(10,240)
\qbezier(10,240)(20,245)(20,260)
\qbezier(20,220)(20,235)(10,240)
\qbezier(10,240)(0,245)(0,260)
\put(0,180){\vector(0,-1){200}}
\qbezier(40,300)(40,285)(30,280)
\qbezier(30,280)(20,275)(20,260)

\put(210,-15){\line(1,0){80}}
\put(210,-15){\line(-2,1){20}}
\put(210,-15){\line(-2,-1){20}}
\put(290,-15){\line(2,1){20}}
\put(290,-15){\line(2,-1){20}}
\put(248,-10){\mbox{$0$}}
\put(210,215){\line(1,0){80}}
\put(210,215){\line(-2,1){20}}
\put(210,215){\line(-2,-1){20}}
\put(290,215){\line(2,1){20}}
\put(290,215){\line(2,-1){20}}
\put(248,220){\mbox{$l$}}
\qbezier(186,210)(182,215)(186,220)
\put(186,220){\vector(1,2){2}}
\put(186,210){\vector(1,-2){2}}
\put(165,212){\mbox{$T_l^2$}}
\put(135,212){\mbox{$\bigoplus_{l}$}}
\put(250,60){\line(0,1){20}}
\put(250,60){\line(-2,-1){20}}
\put(250,60){\line(2,-1){20}}
\put(250,80){\line(-2,1){20}}
\put(250,80){\line(2,1){20}}
\put(254,67){\mbox{$j$}}
\qbezier(234,95)(250,110)(266,95)
\put(236,97){\vector(-1,-1){2}}
\put(264,97){\vector(1,-1){2}}
\put(246,110){\mbox{$T_j^{2k}$}}
\put(250,290){\line(0,1){20}}
\put(250,290){\line(-2,-1){20}}
\put(250,290){\line(2,-1){20}}
\put(250,310){\line(-2,1){20}}
\put(250,310){\line(2,1){20}}
\put(254,297){\mbox{$0$}}
%
%
\put(-12,-25){\mbox{$\bar r$}}
\put(23,-25){\mbox{$r$}}
\put(68,-25){\mbox{$\bar r$}}
\put(103,-25){\mbox{$r$}}
%
%
%
\put(182,-30){\mbox{$\bar r$}}
\put(182,-7){\mbox{$r$}}
\put(315,-7){\mbox{$\bar r$}}
\put(315,-30){\mbox{$r$}}
\put(221,50){\mbox{$\bar r$}}
\put(221,90){\mbox{$r$}}
\put(275,90){\mbox{$\bar r$}}
\put(275,50){\mbox{$r$}}
\put(200,70){\mbox{$\bigoplus_j$}}
\put(182,228){\mbox{$r$}}
\put(182,197){\mbox{$\bar r$}}
\put(315,200){\mbox{$ r$}}
\put(315,228){\mbox{$\bar  r$}}
\put(221,280){\mbox{$\bar r$}}
\put(221,320){\mbox{$r$}}
\put(275,320){\mbox{$\bar r$}}
\put(275,280){\mbox{$r$}}
\put(250,10){\vector(0,1){30}}
\put(250,145){\vector(0,1){40}}
\put(250,245){\vector(0,1){20}}
\put(257,20){\mbox{$S_{j0}$}}
\put(257,160){\mbox{$S_{lj}$}}
\put(257,250){\mbox{$S_{0l}$}}
\end{picture}

Note that in order to have a closed oriented line one should no change
the order in which representation and its conjugate appear in the last two
vertical lines.

The analytical expression is now
\be
\sum_{l,j} S_{0l}\left[\begin{array}{cc} r & \bar r \\ \bar r & r\end{array}\right] \
T_l\big[r,r\big]^2 \
S_{lj}\left[\begin{array}{cc} r & \bar r \\ \bar r & r\end{array}\right] \
T_j\big[r,\bar r\big]^{2k} \
S_{j0}\left[\begin{array}{cc} r & \bar r \\ \bar r & r\end{array}\right] \
\ee
In the case of the fundamental representation $r=[1]=\overline{[1]}$ of $SU(2)$
one can use (\ref{ST2}) and obtain:
\be
ST^2ST^{2k}S \ \stackrel{(\ref{ST2})}{=}
\frac{1}{[2]^2} \left(\begin{array}{ccc} q^{2k-1}(q^2+q^{-2})+q^{-2k}\{q^3\} &&
\Big(q^{2k-1}(q^2+q^{-2})-q^{-2k}\{q\}\Big)\sqrt{[3]} \\ \\
\Big(q^{2k}\{q\}+q^{1-2k}(q^2+q^{-2})\Big)\sqrt{[3]} && q^{2k}\{q^3\}-q^{1-2k}(q^2+q^{-2}) \end{array}\right)
= \nn \\ =
\ \left(\begin{array}{ccc} \ldots && \ldots \\ \\ \ldots &&
-\frac{q^{-2k-2}}{[2]}J_{[1]}^{Tw(k)} \end{array}\right)
\label{STtw2}
\ee

\be
J_{[1]}^{Tw(k)} = \left. 1+ \frac{A^{k+1}\{A^{-k}\}}{\{A\}}\{Aq\}\{A/q\}\right|_{A=q^2}
= \frac{1}{[2]}\Big(-q^{4k+2}\{q^3\}+ q^3(q^2+q^{-2})\Big)
= \nn \\
= -\frac{q^{2k+2}}{[2]}\Big( q^{2k}\{q^3\} - q^{1-2k}(q^2+q^{-2})\Big)
\ee

\end{itemize}

\bigskip

\paragraph{3 caps}

The initial state in this case can be represented as

\begin{picture}(100,120)(-120,-100)
\put(0,0){\vector(0,-1){20}}
\put(20,-20){\vector(0,1){20}}
\qbezier(0,-20)(0,-30)(10,-30)
\qbezier(20,-20)(20,-30)(10,-30)
\put(100,0){\vector(0,-1){20}}
\put(80,-20){\vector(0,1){20}}
\qbezier(80,-20)(80,-30)(90,-30)
\qbezier(100,-20)(100,-30)(90,-30)
\put(160,0){\vector(0,-1){20}}
\put(180,-20){\vector(0,1){20}}
\qbezier(160,-20)(160,-30)(170,-30)
\qbezier(180,-20)(180,-30)(170,-30)
\put(10,-50){\line(1,0){160}}
\put(10,-50){\line(-2,1){20}}
\put(10,-50){\line(-2,-1){20}}
\put(170,-50){\line(2,1){20}}
\put(170,-50){\line(2,-1){20}}
\put(90,-50){\line(0,-1){30}}
\put(90,-80){\line(-2,-1){20}}
\put(90,-80){\line(2,-1){20}}
\put(48,-47){\mbox{$0$}}
\put(128,-47){\mbox{$0$}}
\put(94,-70){\mbox{$0$}}
\end{picture}

This picture shows one of the possible mutual orientations of the three lines,
which is suitable for the following example.
One can now make the chain of modular transformations
(from the above 2-cap examples it should be clear what is the
associated link diagram, but analytical expressions
can be read from the chain of modular transformations):

\begin{picture}(300,200)(-60,-90)
\put(0,50){\line(1,0){80}}
\put(0,50){\line(-2,1){20}}
\put(0,50){\line(-2,-1){20}}
\put(80,50){\line(2,1){20}}
\put(80,50){\line(2,-1){20}}
\put(40,50){\line(0,1){30}}
\put(40,80){\line(-1,2){10}}
\put(40,80){\line(1,2){10}}
\put(18,40){\mbox{$0$}}
\put(58,40){\mbox{$0$}}
\put(44,62){\mbox{$0$}}
\put(210,50){\line(1,0){120}}
\put(210,50){\line(-2,1){20}}
\put(210,50){\line(-2,-1){20}}
\put(330,50){\line(2,1){20}}
\put(330,50){\line(2,-1){20}}
\put(250,50){\line(-1,2){10}}
\put(290,50){\line(1,2){10}}
\put(228,40){\mbox{$0$}}
\put(308,40){\mbox{$0$}}
\put(268,55){\mbox{$j$}}
\put(250,-50){\line(1,0){40}}
\put(250,-50){\line(-2,-1){20}}
\put(250,-50){\line(-2,1){40}}
\put(290,-50){\line(2,-1){20}}
\put(290,-50){\line(2,1){40}}
\put(230,-40){\line(-1,2){10}}
\put(310,-40){\line(1,2){10}}
\put(267,-46){\mbox{$j$}}
\put(237,-40){\mbox{$k$}}
\put(298,-40){\mbox{$l$}}
\put(40,-50){\line(0,-1){20}}
\put(40,-70){\line(-2,-1){20}}
\put(40,-50){\line(-2,1){40}}
\put(40,-70){\line(2,-1){20}}
\put(40,-50){\line(2,1){40}}
\put(20,-40){\line(-1,2){10}}
\put(60,-40){\line(1,2){10}}
\put(27,-40){\mbox{$k$}}
\put(48,-40){\mbox{$l$}}
\put(44,-62){\mbox{$m$}}
\put(120,50){\vector(1,0){50}}
\put(140,55){\mbox{$S_{j0}$}}
\put(270,20){\vector(0,-1){30}}
\put(275,3){\mbox{$S_{k0}\otimes S_{l0}$}}
\put(170,-50){\vector(-1,0){50}}
\put(140,-45){\mbox{$S_{mj}$}}
\end{picture}

Now one can apply $T$ transformations in any of the channels,
moving back and forth along this chain.

The typical analytical expression begins from:
\be
\ldots \
T_m\big[r,\bar r\big]^{a}\
S_{mj}\left[\begin{array}{cc}  k & l  \\ \bar t  & r  \end{array}\right]\
T_k\big[r,\bar r\big]^{b}\ T_l\big[r,\bar r\big]^{c}\
S_{l0}\left[\begin{array}{cc}  \bar r & \bar r  \\ \bar j  & \bar r  \end{array}\right]
S_{k0}\left[\begin{array}{cc}  r & r  \\ \bar r  & \bar j  \end{array}\right]
S_{j0}\left[\begin{array}{cc}  r & \bar r   \\ 0  & 0  \end{array}\right]
\ee
As usual, it is read from right to left, and one can add arbitrary many $S$ transforms and their conjugate of the same type to the left.

Note that the obvious selection rule dictates that $j=r$,
thus actually there is {\it no} sum involving {\it arguments} (not just indices)
of the matrix $S$.
However, such sums can appear after additional applications of $S$.

\subsection{Hikami knot invariants from check-operators}

There is another, alternative construction of ${\cal R}$-matrices, which has a geometric origin and is associated with tetrahedra volume \cite{Kash,Hikami,Hik1,Hik2}. It is basically associated with Chern-Simons theory with a complex gauge group $G_{\IC}$ \cite{Guk}.

\subsubsection{Quantum spectral curve in Chern-Simons theory}\label{sec:KZtoBPZ}

The  Chern-Simons theory on a 3d manifold ${\cal M}$ with a complex gauge group $\mathfrak{SL}(2,\IC)$ is defined by the following action \cite{CS-cplx}
\be
S(\A,\tilde\A)=\frac{t_+}{8\pi}\int\lm_{\cal M}\Tr\left(\A\wedge d\A+\frac{2}{3}\A\wedge\A\wedge \A\right)+\frac{t_-}{8\pi}\int\lm_{\cal M}\Tr\left(\tilde\A\wedge d\tilde\A+\frac{2}{3}\tilde\A\wedge\tilde\A\wedge\tilde\A\right)
\ee
where
\be
t_{\pm}=k\pm i s, \quad k\in \IZ,\quad s\in\IR
\ee
and the path integral runs over both $\A$ and $\bar\A$:
\be
\int D\A D\bar\A\; e^{i S(\A,\bar\A)}
\ee
Here we are going to deal with correlation functions of only fields $\A$. Since we are interested in constructing knot polynomials in this theory, i.e. the Wilson averages, we follow the logic of s.5.1: consider a monodromy of the wave function as a time evolution.
We consider the case when $\cal C$ is represented by a sphere and the gauge group is $\mathfrak{SL}(2,\IC)$. Fixing the representations along the Wilson lines, one can represent the average of the Wilson lines that begin at points $v_i$ and finish at points $x_i$ as the tensor
\be
\rho\left[\left\langle \bigotimes\lm_i \Pexp\int\lm_{v_i}^{x_i}{\cal A}\right\rangle\right]=\bigotimes\lm_i \rho_i(g(x_i)) \bigotimes\lm_j \rho_j(g^{-1}(x_j))\Psi(x_1,\ldots,x_n, v_1,\ldots, v_n)
\ee
The tensor-valued wave function $\Psi$ here is a conformal block of the WZWN theory, which satisfies the Knizhnik-Zamolodchikov equation with respect to the both sets of variables $v_i$ and $x_i$:
\be\label{KZ}
(t_+-2)\p_{x_i}\Psi=\sum\lm_{j\neq i}\frac{\rho_i(\tau^a)\otimes \rho_j(\tau^a)}{x_i-x_j}\Psi
\ee
Let one of the representations be fundamental and denote the corresponding end-point as $z$, while the remaining ones are $x_i$. Then,
\be\begin{split}
(t_+-2)\p_z\Psi=\Phi(z)\Psi,\ \ \ \ \ \ \ \ \ \ \ \
\Phi(z):=\sum\lm_i \frac{\sigma^a\otimes \rho_i(\tau^a)}{z-x_i}\end{split}
\ee
where $\sigma^a$ are the Pauli matrices. The equation for the first component $\Psi_1$ is
\be
\hbar \p_z^2\Psi_1= \Phi_{11} \p_z \log \left({\Phi_{11}\over\Phi_{12}}\right)\cdot \Psi_1 + \hbar \p_z \log \Phi_{12}\cdot \p_z\Psi_1 +{1\over\hbar} \Big(\Phi^2\Big)_{11}\cdot\Psi_1
\ee
with $\hbar=t_+-2$. In the quasiclassical limit, only the last term at the r.h.s. of this equation survives so that one finally obtains
\be\label{sc}
\hbar \p_z^2\Psi_1=\sum\lm_i\frac{c_2(\rho_i)}{(z-x_i)^2}\Psi_1+\sum\lm_i\frac{1}{z-x_i}\underbrace{\sum\lm_{j\neq i}\frac{\rho_i(\tau^a)\otimes \rho_j(\tau^a)}{x_i-x_j}}_{\hbar\p_{x_i}}\Psi_1
\ee
with $c_2(\rho)$ being the second Casimir element.
This is the spectral curve equation of the form (\ref{keq}), with the potential parameterized by $z_i$ and, hence, the check-operator acts on $z_i$. The potential itself can be restored from comparing this equation and (\ref{keq}).
This similarity of (\ref{sc}) and (\ref{keq}) allows one to make ``$\beta$-ensemble interpretations'' in Chern-Simons theory.
The derivative $\p_z$ is definitely replaced in the quasiclassical approximation with the spectral parameter $\lambda$.

\subsubsection{Verlinde operators}

Consider an operator that acts on the Hilbert space in Chern-Simons theory
\be
O_{\gamma}^{(R)}:\; {\cal H}_{CS}\rightarrow {\cal H}_{CS}
\ee
and define from the Knizhnik-Zamolodchikov equation (\ref{KZ}) an ``evolution'' operator that moves points of the wave function, or conformal block from their initial positions $z_i$ (the initial moment of the evolution, $t=0$) to  the final positions after monodromy transformation made $v_i$ (the final moment of the evolution, $t=T$):
\be
{\cal U}({\rm Knot}):=\bigotimes\lm_{i} \Pexp \int\lm_{i {\rm th \; strand}} A_i d\zeta_i=\bigotimes\lm_{i} \Pexp \int_{z_i}^{v_i} d\zeta_i\;\frac{1}{\hbar}\sum\lm_{j\neq i}\frac{\rho_i(\tau^a)\otimes \rho_j(\tau^a)}{\zeta_i-\zeta_j}
\ee
so that one can generate knots invariants evaluating either the trace of this operator in $\bigotimes\lm_i R_i$ or its projection. In the Heisenberg representation
\be
{\cal U}({\rm Knot})^{-1}O_{\gamma}^{(R)}(0)\;{\cal U}({\rm Knot})=O_{\gamma}^{(R)}(T)
\ee
and if one knows both $O_{\gamma}^{(R)}(0)$ and $O_{\gamma}^{(R)}(T)$, it is possible to evaluate ${\cal U}({\rm Knot})$ from the equation
\be
\boxed{O_{\gamma}^{(R)}(0)\; {\cal U}({\rm Knot})={\cal U}({\rm Knot})\; O_{\gamma}^{(R)}(T)}\label{Heis}
\ee
Let us introduce a set of ``Verlinde operators'' acting on the space of conformal blocks ${\cal H}_{CS}$ as the monodromy trace (straight analogs of $W_R$ from s.\ref{sec:Reidemeister}):
\be
O_{\gamma}^{(R)}:=\Tr_R \; \Pexp \oint\lm_\gamma d\zeta\;\frac{1}{\hbar}\sum\lm_i \frac{\vec{\tau}\otimes\vec{\tau}_i}{\zeta-z_i}
\ee
They can be rewritten following ``the $\beta$-ensemble interpretation'' in terms of check-operators as the trace of monodromy:
\be\label{O}
O_{\gamma}^{(R)}=\Tr_R e^{{1\over\hbar}\oint\lm_{\gamma}\check{\nabla}}
\ee
while the counterparts of monodromy itself are the (Fock-Goncharov) cluster coordinates
\be\label{cc}
\X_{\gamma}:=e^{{1\over\hbar}\oint\lm_{\gamma}\check{\nabla}}
\ee
which form a Heisenberg algebra:
\be\label{HA}
\X_{\gamma}\X_{\gamma'}=q^{\langle\gamma,\gamma'\rangle}\X_{\gamma+\gamma'}
\ee
Then, one can solve equation (\ref{Heis}) in terms of this algebra:
\be
\boxed{{\cal U}({\rm Knot})=f(\X_{\gamma})}
\ee

\bigskip

This algebra admits a realization in the space of conformal blocks \cite{GMM} with manifest realization
$\X_A=e^a$, $\X_B=q^{\p_a}$ so that the evolution operator is realized as ${\cal U}({\rm Knot})=f(e^a, q^{\p_a})$ and reduces to a modular transformation of the conformal block in terms of $S$- and $T$-matrices/kernels.

\subsubsection{Knots and flips}

\paragraph{Quasiclassical limit.}
As we described, one can associate with each WZWN conformal block the corresponding Knizhnik-Zamolodchikov equation (and its derivative (\ref{sc})), and with this later a WKB network. When the points of the conformal block are subject to monodromy transformations, this Chern-Simons evolution can be described by reconstructions of the WKB network by a series of flips (mutations).  In terms of the Heisenberg algebra (\ref{HA}) associated with the Stokes lines $\gamma$'s, we {\bf reinterpret} flips as an action of some evolution operators $u$ on $\X_\gamma$, and the {\bf discretized} smooth evolution is now
\be
{\cal U}({\rm Knot})\leadsto \prod\lm_{\gamma \in {\rm flips}}  u_{\gamma}(X)
\ee

Therefore, one can consider the spectral curve that emerges in the ``quasiclassical'' limit $\hbar\rightarrow 0$, (\ref{sc}): $\check\nabla^2(z)-T(z)=0$
and the Stokes lines $\boxed{\mathop{\rm Im} \hbar^{-1}\check\nabla=0}$ (see (\ref{BPS})) so that we are able to present quasiclassical expressions for {\bf the operators}:
\be
O_{\gamma}^{(R)}\sim \sum\lm_{\rm sheets} e^{\oint\lm_{\gamma}\check{\nabla}} +\mbox{\bf Stokes detours}
\ee
A single flip along the edge $\gamma$ is calculated as in s.2.2 and is equal (see eq.(\ref{flip}))
\be
{\rm Flip}_{\gamma}(\X_{\gamma'})=\left\{\begin{array}{ll}
	\X_{\gamma'}, & \langle\gamma,\gamma'\rangle=0\\
	\X_{\gamma'}(1+\X_{\gamma}), & \langle\gamma,\gamma'\rangle=1	
\end{array}\right.
\ee

\paragraph{Quantization.}
Similarly to s.4.1, using technique presented in \cite{GLM} one can calculate the quantum flip. The result is
\be
{\rm Flip}_{\gamma}(\X_{\gamma'})=\left\{\begin{array}{ll}
	\X_{\gamma'}, & \langle\gamma,\gamma'\rangle=0\\
\X_{\gamma'}\prod_{a=1}^{|\langle\gamma,\gamma'\rangle|}\Big(1+
q^{2a-1}\X_{\gamma}^{\langle\gamma,\gamma'\rangle}\Big)^{\langle\gamma,\gamma'\rangle}, & |\langle\gamma,\gamma'\rangle|=1
\end{array}\right.
\ee
This flip is described by the adjoint action of $u_\gamma$:
\be
\X_{\gamma'} u_{\gamma}(\X)=u_{\gamma}(\X){\rm Flip}_{\gamma}(\X_{\gamma'})
\ee
which means that
\be\label{dl}
\boxed{
u_{\gamma}(\X)\sim \Phi(\log \X_{\gamma})}
\ee
where quantum dilogarithm is defined as \cite{Fad}
\be\label{qd}
\Phi(z|\tau)=\exp\left(-\frac{1}{4}\int_{{\bf R}+i0}\frac{dw}{w}\frac{e^{-2 iw z}}{\sinh b^{-1}w \cdot\sinh b w}\right)
\ee

\subsubsection{Hikami invariant as a KS invariant }

Thus, one is able to rewrite the evolution operator ${\cal U}(\hbox{Knot})$ as a product of ${\cal R}$-matrices each of them, in its term, being a product of mutations:
\be\label{eq:R-mut}
{\cal R}=:\prod\lm_{i}u_{\gamma_i}:
\ee
Having the manifest expression for $u_{\gamma}$ in terms of quantum dilogarithm (\ref{qd}), one can construct a manifest representation for the ${\cal R}$-matrix. This can be done either by using our manifest realization of the cluster coordinates (\ref{cc}), or in a more formal way \cite{Hik2}, the answer for the ${\cal R}$-matrix being a product of ratios of quantum dilogarithms. In order to obtain the knot polynomial one still has to calculate the trace of a product of ${\cal R}$-matrices.

Afterwards the $R$-matrix can be rewritten in terms of tensor categories after substitution of values for cluster coordinates in terms of tensor categories.
From this point of view $\mathcal{V}ir_c$ and $\mathcal{U}_q(sl_2)$ are equivalent tensor categories \cite{PT}\footnote{Indeed this equivalence can be explicitly demonstrated \cite{GMM}, and on general grounds it is a consequence of a mythical mirror symmetry in a mythical $\mathcal{V}ir_{q,t}$-tensor category.}.

\paragraph{Quasiclassical limit \cite{Hik1}.}
As was demonstrated in \cite{Hik1,Hik2}, the ${\cal R}$-matrix can be associated with an ideal hyperbolic octahedron. This is not surprising since the quantum dilogarithm (\ref{qd}) in the quasiclassical limit ($q\rightarrow 1$) is related to the hyperbolic volume of ideal tetrahedron $\Delta$ so that the ${\cal R}$-matrix has the asymptotic form
\be
{\cal R}\sim e^{\frac{1}{\hbar^2}\sum\lm_i\Delta_{i}}
\ee

\subsection{Stokes phenomenon in conformal blocks}
\subsubsection{WKB approximation}
As we have seen the spectral curve for a braid of $n$ strands placed at $x_i$ reads (\ref{sc})
\be
\lambda^2=\sum\lm_{i=1}^n\left(\frac{c_2(\rho_i)}{(z-x_i)^2}+\frac{u_i}{z-x_i}\right)dz^2
\ee
The web of WKB lines is constructed as trajectories of solutions to equation (\ref{BPS})
\be
\mathop{\rm Im}\; \hbar^{-1}\lambda=0
\ee
Here we present the evolution of the WKB lines for six strands associated to the premutation of two middle strands in order to mimic the action of the ${\cal R}$-matrix. The blue dots mark positions of zeroes of the discriminant, while the red ones mark positions of the strands (singularities of the curve).

\begin{figure}[htbp]
	\begin{center}
		\includegraphics[scale=0.5]{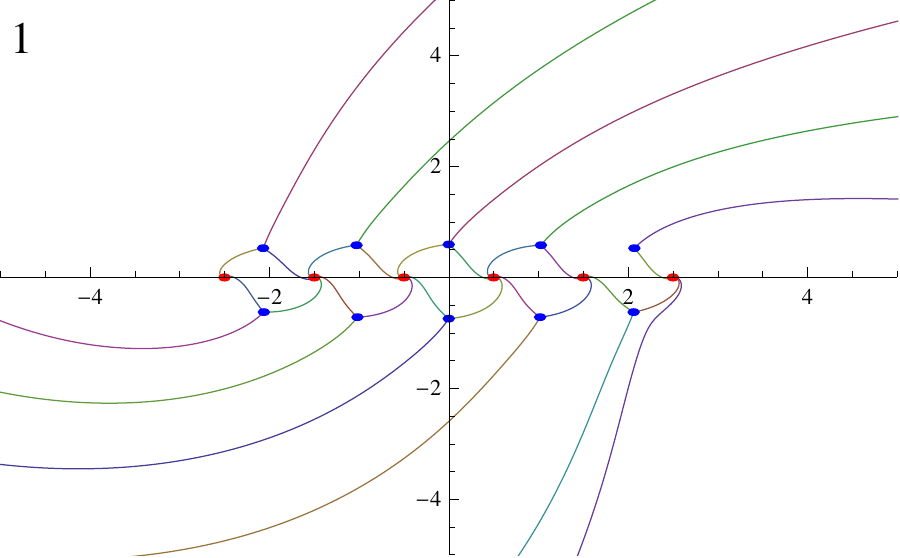}\;\;\;\;\;\;\;\;\includegraphics[scale=0.5]{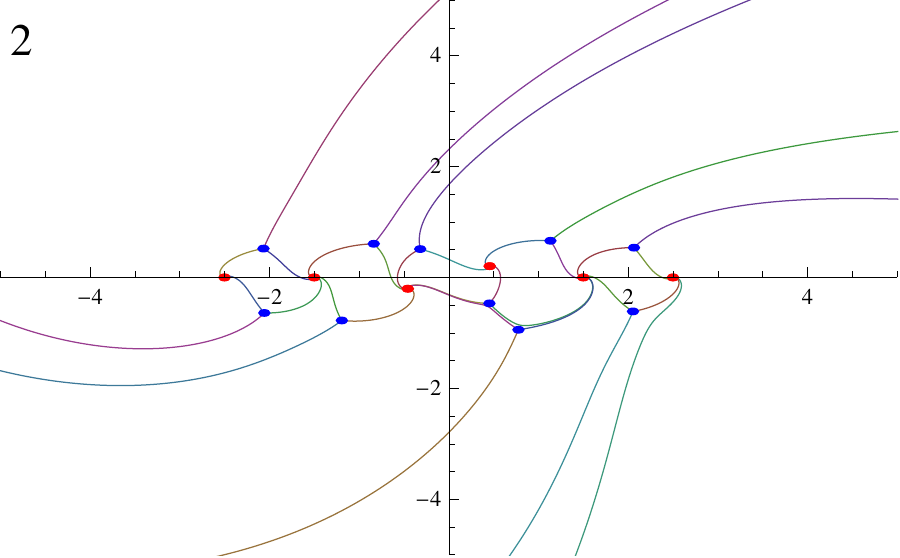}
		
		\bigskip
		
		\includegraphics[scale=0.5]{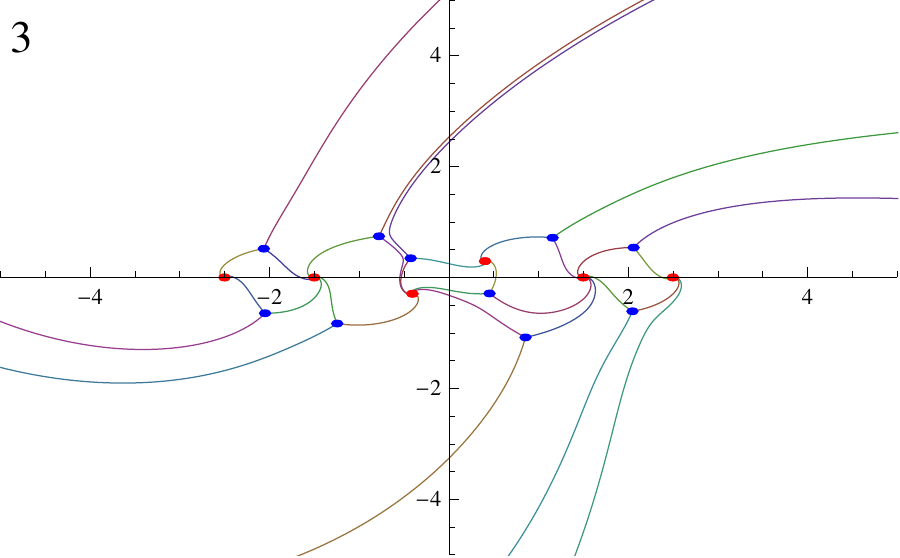}\;\;\;\;\;\;\;\;\includegraphics[scale=0.5]{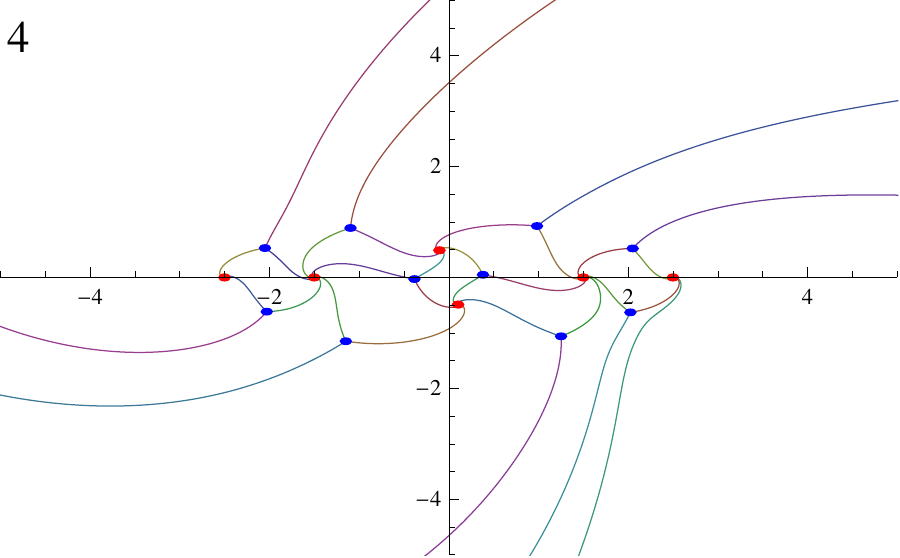}
		
		\bigskip
		
		\includegraphics[scale=0.5]{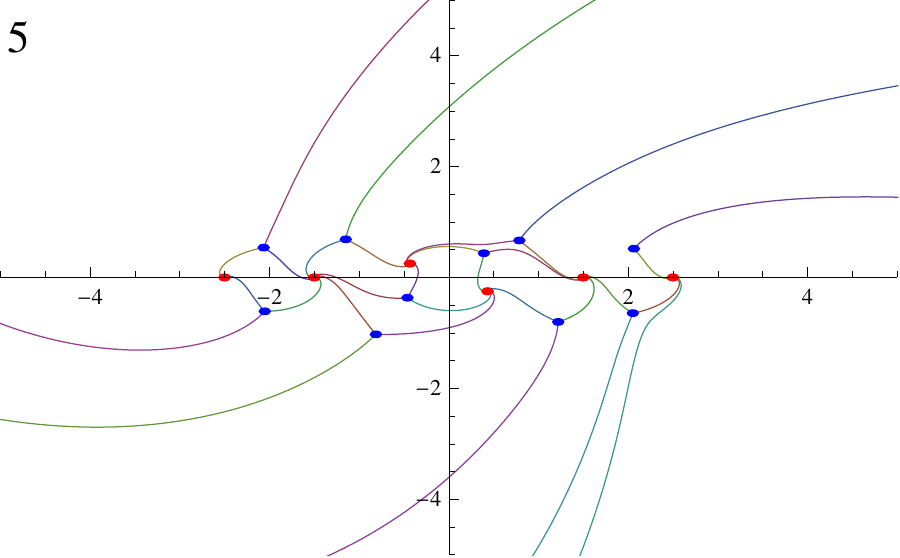}\;\;\;\;\;\;\;\;\includegraphics[scale=0.5]{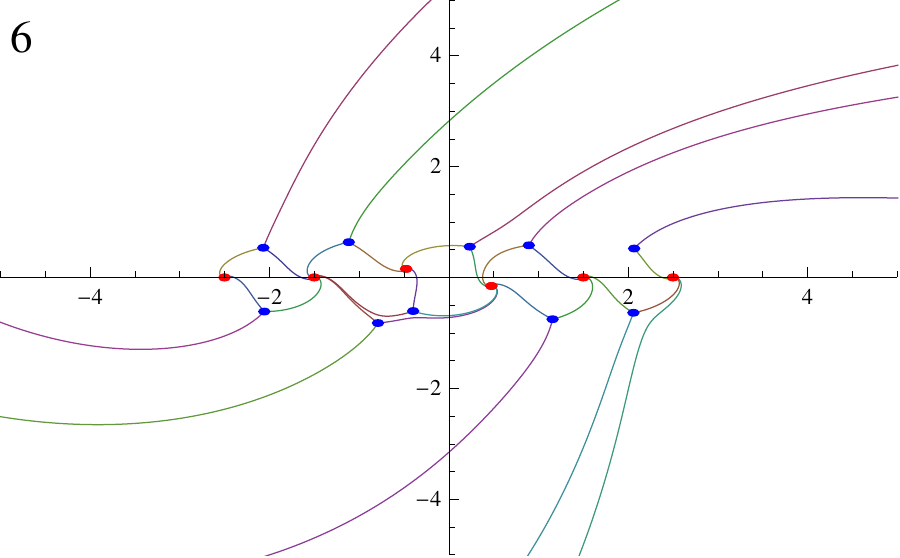}
		\caption{Evolution of WKB lines.\label{fig:WKB}}
	\end{center}
\end{figure}	

\begin{figure}[htbp]
	\begin{center}
		\includegraphics[scale=0.5]{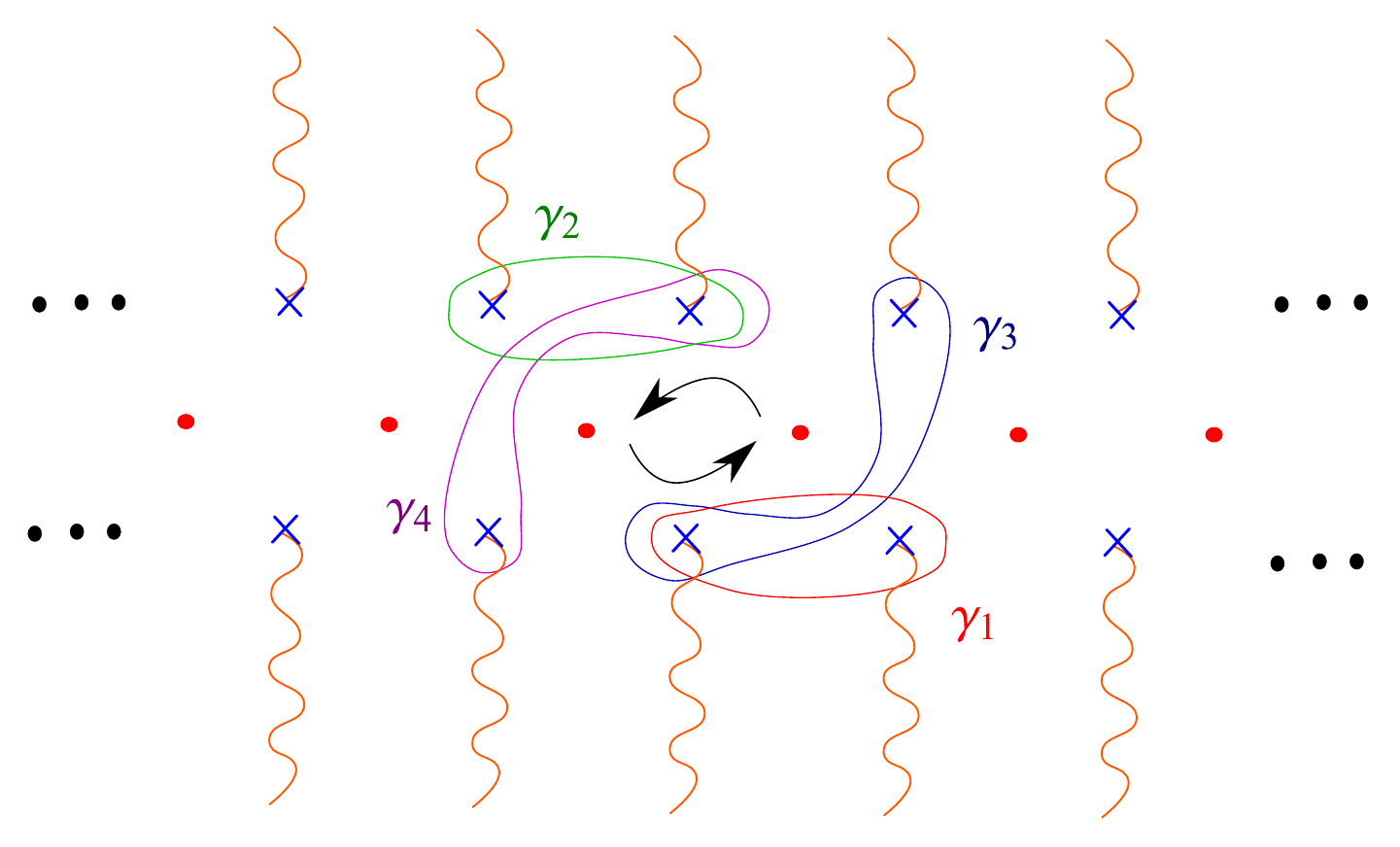}
		\caption{Cycles associated to the flips corresponding to the permutation of two strands.\label{fig:R-mat}}
	\end{center}
\end{figure}

It is simple to observe four simple flips associated to cycles $\color{red} \gamma_1$, $\color{green}\gamma_2$, $\color{NavyBlue}\gamma_3$, $\color{Purple}\gamma_4$ in the order as they are mentioned as it is depicted in Fig.\ref{fig:R-mat}.

Hence, the action of the ${\cal R}$-matrix is mimiced by the following operator (cf. with \cite[eq.(3.15)]{Hik2})
\be
\boxed{{\cal R}\sim\Phi(\X_{\color{Purple}\gamma_4})\Phi(\X_{\color{NavyBlue}\gamma_3})\Phi(\X_{\color{green}\gamma_2})\Phi(\X_{\color{red}\gamma_1})}
\ee

\begin{figure}[htbp]
	\begin{center}
		\includegraphics[scale=1.0]{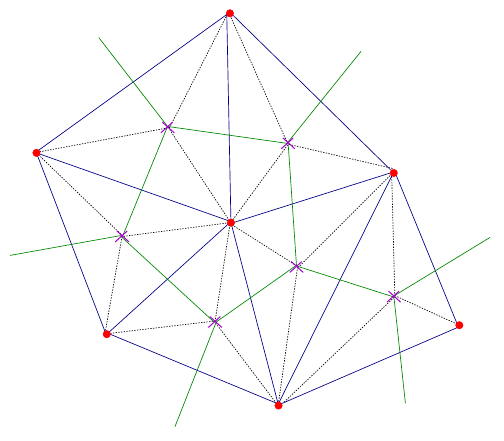}
		\caption{Triangulation.\label{fig:Fock_Gonch-Kash}}
	\end{center}
\end{figure}

To clarify the relation to what is discussed in \cite{Hik2}, we should mention the relation between the Fock-Goncharov and the Kashaev coordinates on triangulations. Consider a triangulation of a punctured Riemann surface depicted in fig.\ref{fig:Fock_Gonch-Kash}. Here the singularities are marked as red dots, branching points are marked as purple crosses, and the WKB lines are dashed lines. Then, we restore the triangulation edges as it is marked by the blue lines. One can associate the Kashaev coordinates to these edges. Notice that the Fock-Goncharov coordinates are associated with the cycles that are projected to the green lines connecting the branching points, the centers of triangles. In other words, the Fock-Goncharov coordinates are associated with the graph dual to the triangulation and are dual to the Kashaev coordinates correspondingly.

The antisymmetric matrix $B_{ij}$ associated with a quiver (see \cite[eq.(2.1)]{Hik2}) represents an intersection pairing matrix associated to the corresponding cycles.

In the next fig.\ref{fig:triang}, we present the WKB triangualtion of the spectral curve under consideration and flips of its edges associated with $\color{red} \gamma_1$, $\color{green}\gamma_2$, $\color{NavyBlue}\gamma_3$, and $\color{Purple}\gamma_4$ respectively marked by the corresponding colors (the initial edge is marked by the solid line, the mutated edge is marked by the dashed line.) Comparing this triangulation to that presented in \cite[Figure 2]{Hik2}, we notice that all the horizontal edges are flipped from the very beginning in that paper as compared with the present ones, and the top and bottom tips are not glued together.

\begin{figure}[htbp]
	\begin{center}
		\includegraphics[scale=0.7]{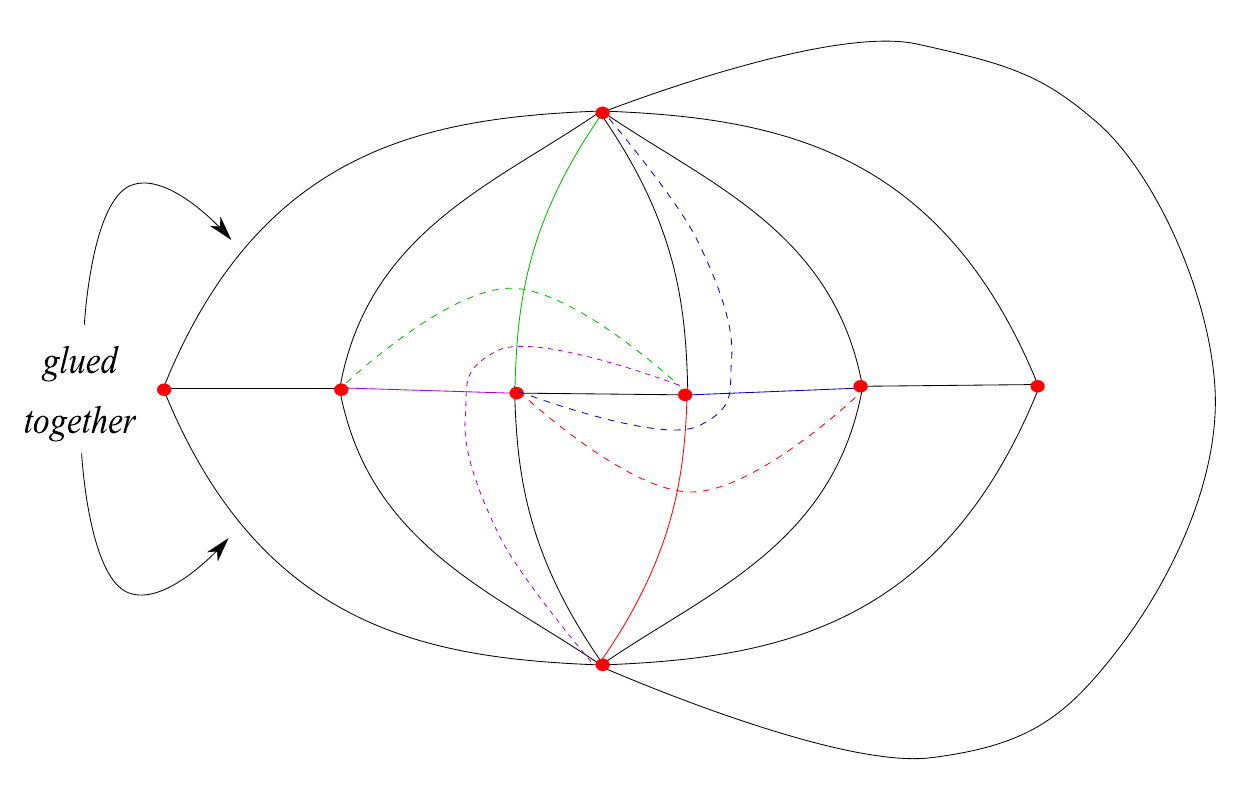}
		\caption{Flips of the triangulation associated with the conformal block spectral curve.\label{fig:triang}}
	\end{center}
\end{figure}

\section{Conclusion and Discussion}

In this short review we tried to give a simple intuitive description for various new interesting phenomena described recently in the literature. The story includes such issues as quantum spectral curves, Teichm\"uller theory, cluster varieties, moduli space of flat connections, different theories produced by M5-brane compactification, etc...

Many of these subjects are related to a 2d Coulomb gas system ($\beta$-ensemble), maybe via different chains of dualities or correspondences. The usual technique to derive correlators in the $\beta$-ensemble system, the topological recursion gives rise to an infinite chain of linked equations: the loop equations. We argued that the loop equations arising within this approach are much similar to those arising within the WKB approach to solving differential equations. The major modification is that in $\beta$-ensembles there are two deformation coefficients: $g$, a string coupling constant, and $\beta$. This can be reinterpreted in such a way that, to the usual Plank constant $\hbar$ controlling WKB expansion of, say, Schr\"odinger equation one should add another deformation parameter $\hbar'$ that will control commutation relations of eikonals. Thus, the eikonals become non-commutative operators acting on some modified Hilbert space, where the $\beta$-ensemble partition function behaves as a wave function on the moduli space.

As we discussed, the main characters of this construction are: the flat connection on Riemann surface, the holonomy of this connection, the spectral parameter or eikonal, the cluster coordinates arising as holonomies of this eikonal, and the Hilbert space associated with the partition faction playing a role of the wave function. As well, many theories are tightly related to some dualities. Here we present various avatars of the same notions emerging in different theories. Some of the avatars have been so far not discussed in the literature, hence, we leave the corresponding boxes blank:

\bigskip

\begin{tabular}{|p{2.3cm}|p{2cm}|p{2cm}|p{2cm}|p{2cm}|p{2cm}|p{2cm}|}
	\hline
	Theory &Hilbert space& Flat connection & holonomy & spectral parameter & cluster coordinate & duality \\
	\hline
	\hline
	2d CFT &conf. block& $T(z)$ & Verlinde operator\cite{Ga,Go} & chiral current\cite{Chekhov} & Y-functions & fusion rules \\
	\hline
	$\beta$-ensembles & part. func.&  &  & resolvent &  & S-duality \\
	\hline
	top. strings &part. func.&  &  & $U(1)$-Chern-Simons & Wilson loops &  \\
	\hline
	Knizhnik-Zamolodchikov-Bernard &conf. block, CS wave function& $\mathfrak{g}$-connection  & Wilson loops &  &  & Reidemeister moves \\
	\hline
	Quantum Hitchin &Landau-Ginsburg wave func.& Lax & Lax holonomies &  &  & spectral dualities MMZZ\\
	\hline
	q-Teichm\"uhler &wave func.&spin connection &spin conn. holonomies & geodesics & Fock-Goncharov, Kashaev & Moore-Seiberg groupoid\\
	\hline
	Nekrasov &part. func.& 2d defect UV & 1d defect UV & 2d defect IR & 1d defect IR & S-duality\\
	\hline
\end{tabular}

\bigskip

To complete this table with references we present the following list and web of correspondences between different theories and references:
\begin{enumerate}
	\item 2d CFTs: \cite{AGT,AGT1}
	\item CS/topological strings: \cite{Cecotti-Vafa}
	\item CS/KZB-equations: \cite{Moore-Seiberg}
	\item $\beta$-ensembles: \cite{MM}
	\item Quantum integrable systems (q-Hitchin, Painlev\'e): \cite{Iorgov,MMZZ}
	\item $R$-twisted $tt^*$-equations: \cite{Cecotti-Vafa, GMN,GMN2}
	\item Quantum Teichm\"uhler theory: \cite{qT}
	\item ${\cal N}=2$ 4d theory in $\Omega$-background: \cite{AGT,AGT1}
	\item Reps of quantum groups: \cite{PT}
\end{enumerate}
$$
\xymatrix{
	\mbox{M-theory}\ar@{<->}[d]^{\cite{Cecotti-Vafa}}\ar@/_5pc/@{<->}[dd] 
	&&\\
	\mbox{CS/topological strings}\ar@{<->}[r]^{\cite{Cecotti-Vafa}}\ar@{<->}[dr]^{\cite{Aganagic}}&\mbox{$R$-twisted $tt^*$-equation} \ar@{<->}[r]^{\cite{Gaiotto}}& \mbox{q-Hitchtin (Peinlev\'e)}\ar@{<->}[d]_{\cite{Olshanetsky}} \ar@/^8pc/@{<->}[ldd]^{\cite{Iorgov}}\\
	\mbox{(2,0) 6d theory}\ar@{<->}[dr]^{\cite{AGT,AGT1}}\ar@{<->}[d]^{\cite{AGT,AGT1}}&\mbox{$\beta$-ensembles}\ar@{<->}[d]^{\cite{MMS,mamoCFT,GMM1,GMM,Nem}}\ar@{<->}[ru]^{\cite{MMZZ}} & \mbox{CS/KZB-equations} \ar@{<->}[d]\\
	\mbox{${\cal N}=2$ in $\Omega$-background}\ar@{<->}[r]^{\cite{AGT,AGT1}} & \mbox{2d CFT}\ar@{<->}[r]^{\cite{PT}}\ar@{<->}[ru]\ar@{<->}[d]^{\cite{PT}} & \mbox{reps of q-groups}\\
	 & \mbox{Quantum Teichm\"uhler}\ar@{<->}[ru]_{\cite{PT}} & \\
}
$$

In this note we only tried to touch a hip of this iceberg by applying the described framework, in particular, to constructing the Hikami invariants for knots. Nevertheless, the framework seems to be general enough to be extended to include a {\it third} deformation allowing to include quantum W-algebras, 5d SYM theory and desirably superpolynomial invariants for knots.

\section*{Appendix}
\def\theequation{A.\arabic{equation}}

In this Appendix we demonstrate as manifest expressions for the duality matrices can be obtained within the approach due to B.Ponsot and J.Teschner \cite{PT}. We start with a simpler case of $q=e^{\pi i b^2}\rightarrow 1$, which corresponds to $b\rightarrow 0$, $c\rightarrow \infty$ in the conformal theory. The answers for the generic $q$ are obtained just replacing all hypergeometric functions for the $q$-hypergeometric functions, and the $\Gamma$-functions for the $q$-$\Gamma$-functions.

\subsection*{A1. A special case of $q\rightarrow 1$}
\subsubsection*{Preliminaries}
The Ponsot-Teschner approach is made explicitly invariant with respect to $b\leftrightarrow b^{-1}$, so in this limit one can not use already obtained formulas directly. However, one can apply the their framework. The crucial point is that in the limit $c\rightarrow\infty$ the conformal blocks become ``conformal blocks'' of $sl_2$ (see e.g. \cite{Zam,large_c})
\be
\Psi_{\Delta}(x)=B_{\Delta}\left[\begin{array}{cc}
	\Delta_2 & \Delta_3 \\
	\Delta_1 & \Delta_4 \\
\end{array}\right]\rightarrow x^{\Delta-\Delta_1-\Delta_2}{}_2 F_1 \left[\begin{array}{c}
\Delta+\Delta_2-\Delta_1,\; \Delta+\Delta_3-\Delta_4\\
2\Delta
\end{array}\right](x)
\ee
It is an eigenfunction of the operator
\be\label{oper}
\left[\left(x\frac{d}{dx}+\Delta_1+\Delta_2-\frac{1}{2}\right)^2-x\left(x\frac{d}{dx}+\Delta_1+\Delta_2+\Delta_3-\Delta_4\right)\left(x\frac{d}{dx}+2\Delta_1\right)\right]\Psi_{\Delta}(x)=\left(\Delta-\frac{1}{2}\right)^2\Psi_{\Delta}(x)
\ee
This operator is {\bf self-adjoint} with respect to the measure
\be
d\mu(x)=x^{2(\Delta_1+\Delta_2-1)}(x-1)^{\Delta_1-\Delta_2+\Delta_3-\Delta_4}
\ee
Thus, one can calculate the modular kernel as
\be
S_{\Delta\Delta'}\left[\begin{array}{cc}
\Delta_2 & \Delta_3\\
\Delta_1 & \Delta_4\\
\end{array}\right]=\int d\mu(x)\; \Psi^{(s)}_{\Delta}(x)\Psi^{(t)}_{\Delta'}(1-x)
\ee
\subsubsection*{Finite-dimensional representations}
Finite-dimensional representations are described by half-integer spins and correspond to degenerate fields, their dimensions being enumerated by the Kac determinant zeroes, (\ref{Kac})
\be
\Delta([s])=\Delta_{1,s+1}=\frac{b^2 s^2}{4}-\frac{s}{2}\rightarrow-\frac{s}{2}
\ee
For the degenerate fields the conformal blocks are just finite polynomials
\be
\Psi_{j/2}(x)=B_{[j]}\left[\begin{array}{cc}
	[s] & [s] \\
	\left[s\right] & 	\left[s\right] \\
\end{array}
\right] \rightarrow x^{s-\frac{j}{2}}{}_2 F_1 \left[\begin{array}{c}
-\frac{j}{2},\; -\frac{j}{2}\\
-j
\end{array}\right](x),\quad j/2=0,\;1,\;2,\ldots,\; s
\ee
In this case, the modular kernel is just a finite matrix, since
\be\label{nS}
\Psi_{j/2}(x)=\sum\lm_{j'/2=0}^{s}S_{[2j][2j']}\left[\begin{array}{cc}
	[s] & [s] \\
	\left[s\right] & 	\left[s\right] \\
\end{array}\right] \Psi_{j'/2}(1-x)
\ee
For instance,
\be
S_{[j/2][j'/2]}\left[\begin{array}{cc}
	[1] & [1] \\
	\left[1\right] & 	\left[1\right] \\
\end{array}\right]=\left(
\begin{array}{cc}
	-\frac{1}{2} & 1 \\
	\frac{3}{4} & \frac{1}{2} \\
\end{array}
\right)\\
S_{[j/2][j'/2]}\left[\begin{array}{cc}
	[5] & [5] \\
	\left[5\right] & 	\left[5\right] \\
\end{array}\right]=\left(
\begin{array}{cccccc}
	-\frac{1}{6} & \frac{5}{7} & -\frac{25}{14} & \frac{25}{9} & -\frac{5}{2} & 1 \\
	\frac{7}{60} & -\frac{31}{70} & \frac{23}{28} & -\frac{11}{18} & -\frac{1}{4} & \frac{1}{2} \\
	-\frac{7}{90} & \frac{23}{105} & -\frac{1}{12} & -\frac{29}{54} & \frac{7}{12} & \frac{1}{6} \\
	\frac{7}{100} & -\frac{33}{350} & -\frac{87}{280} & \frac{79}{180} & \frac{19}{40} & \frac{1}{20} \\
	-\frac{1}{10} & -\frac{3}{49} & \frac{15}{28} & \frac{95}{126} & \frac{1}{4} & \frac{1}{70} \\
	\frac{11}{36} & \frac{275}{294} & \frac{1375}{1176} & \frac{1375}{2268} & \frac{55}{504} & \frac{1}{252} \\
\end{array}
\right)
\ee
The other matrix $T$ in this case is just $T=\mathop{\rm diag}(1,-1,1,-1,\ldots)$.
In this case, the measure is
\be
d\mu(x)=x^{-2(s+1)}
\ee
and the orthogonality condition for $\Psi_{j}$ is
\be
\left\langle\Psi_{j/2},\Psi_{j'/2}\right\rangle=\int\lm_{1}^{\infty} dx\; x^{j/2+j'/2-2}\left({}_2 F_1 \left[\begin{array}{c}
	-\frac{j}{2},\; -\frac{j}{2}\\
	-j
\end{array}\right](x)\right)\left({}_2 F_1 \left[\begin{array}{c}
-\frac{j'}{2},\; -\frac{j'}{2}\\
-j'
\end{array}\right](x)\right)=\delta_{j,j'}\frac{(2j)!^4}{(4j)!(4j+1)!}
\ee
Particular values of matrix elements are
\be
S_{[0][j/2]}\left[\begin{array}{cc}
	[s] & [s] \\
	\left[s\right] & 	\left[s\right] \\
\end{array}\right]=\frac{1}{||\Psi_{j/2}||^2}S_{[2j][0]}\left[\begin{array}{cc}
[s] & [s] \\
\left[s\right] & 	\left[s\right] \\
\end{array}\right]\\
\begin{split}
S_{[j/2][0]}\left[\begin{array}{cc}
	[s] & [s] \\
	\left[s\right] & 	\left[s\right] \\
\end{array}\right]=\int\lm_{1}^{\infty}dx\; (1-x)^s x^{-s-j/2-2}{}_2F_1 \left[\begin{array}{c}
-\frac{j}{2},\; -\frac{j}{2}\\
-j
\end{array}\right](x)=\\
=(-1)^s\sum\lm_{n=0}^{\infty}\frac{\pi(n-j/2)}{\sin\pi(n-j/2)}\frac{\Gamma(1-j/2+n)}{\Gamma(1-j+n)\Gamma(s+j/2+2-n)n!}
\end{split}
\ee

\subsubsection*{Explicit calculations}

In (\ref{nS}) we consider the normalized matrix $S$
\be
\boxed{S_{kk'}=(-1)^{k+k'-s}\; \frac{\int\lm_1^{\infty}dx \; x^{k-k'-s-2}\; {}_2F_1\left[\begin{array}{c}
		-k\; -k\\
		-2k\\
	\end{array}\right]\left(x^{-1}\right)\; {}_2F_1\left[\begin{array}{c}
	-k'\; -k'\\
	-2k'\\
\end{array}\right](x)}{\int\lm_1^{\infty}dx \; x^{-2(k'+1)} {}_2F_1\left[\begin{array}{c}
		-k'\; -k'\\
		-2k'\\
	\end{array}\right](x)^2},\quad k,k'=0,\ldots,s}
\ee
The normalizing multiplier is independent of $s$, thus, one can evaluate it, for instance, using the OEIS \cite{OEIS}:
\be
\int\lm_1^{\infty}dx \; x^{-2(k'+1)} {}_2F_1\left[\begin{array}{c}
	-k'\; -k'\\
	-2k'\\
\end{array}\right](x)^2=\frac{k!^4}{(2k)!(2k+1)!}=\frac{\Gamma(k+1)^4}{\Gamma(2k+1)\Gamma(2k+2)}
\ee
Now the problem of constructing the matrix $S$ is reduced to evaluating the
\emph{unnormalized} integrals
\be
\tilde S_{kk'}=\int\lm_1^{\infty}dx \; x^{k-k'-s-2}\; {}_2F_1\left[\begin{array}{c}
			-k\; -k\\
			-2k\\
		\end{array}\right]\left(x^{-1}\right)\; {}_2F_1\left[\begin{array}{c}
		-k'\; -k'\\
		-2k'\\
	\end{array}\right](x),\quad k,k'=0,\ldots,s
\ee
In this case, for instance, at $s=1$
\be
\tilde S=\left(
{\renewcommand{\arraystretch}{2.3}\begin{array}{rr}
	\dfrac{1}{2} & \dfrac{1}{12} \\
	\dfrac{3}{4} & -\dfrac{1}{24} \\
\end{array}}
\right)
\ee
We use the Barnes' integral representation for the hypergeometric function
\be
{}_2F_1\left[\begin{array}{c}
	\alpha\;\beta\\
	\gamma\\
\end{array}\right](z)=\frac{\Gamma(\gamma)}{\Gamma(\alpha)\Gamma(\beta)}\frac{1}{2\pi i}\int\lm_{-i\infty}^{i\infty}ds\;\frac{\Gamma(\alpha+s)\Gamma(\beta+s)\Gamma(-s)}{\Gamma(\gamma+s)}(-z)^s
\ee
Thus
\be
\tilde S_{kk'}=\frac{\Gamma(-2k)\Gamma(-2k')}{\Gamma(-k)^2\Gamma(-k')^2}\left(\frac{1}{2\pi i}\right)^2\int\lm_{-i\infty}^{i\infty}dt\;\frac{\Gamma(t-k)^2\Gamma(-t)}{\Gamma(t-2k)}\int\lm_{-i\infty}^{i\infty}dt'\;\frac{\Gamma(t'-k')^2\Gamma(-t')}{\Gamma(t'-2k')}\underbrace{\int\lm_1^{\infty} dx \; x^{k-k'-s-2}\left(-\frac{1}{x}\right)^t (-x)^{t'}}_{\frac{(-1)^{t+t'}}{k-k'+t'-t-s-1}}
\ee
The last term gives a pole so one of the integrals can be calculated
\be
\tilde S_{kk'}=(-1)^{k-k'-s+1}\frac{\Gamma(-2k)\Gamma(-2k')}{\Gamma(-k)^2\Gamma(-k')^2}\int\lm_{-i\infty}^{i\infty}\frac{dt}{2\pi i}\frac{\Gamma(t-k'-s-1)^2\Gamma(-t-k+k'+s+1)}{\Gamma(t-k-k'-s-1)}\frac{\Gamma(t-k')^2\boxed{\Gamma(-t)}}{\Gamma(t-2k')}
\ee
The boxed term gives the poles contributing to the answer. For example,
\be
\tilde S_{11}=\frac{\Gamma(-2)^2}{\Gamma(-1)^4}\left(\frac{\Gamma(-3)^2\Gamma(2)}{\Gamma(-4)}\frac{\Gamma(-1)^2}{\Gamma(-2)}-\frac{\Gamma(-2)^2\Gamma(1)}{\Gamma(-3)}\frac{\Gamma(0)^2}{\Gamma(-1)}\right)=-\frac{1}{24}
\ee
Similarly, one can calculate
\be
\tilde S_{k0}=(-1)^{k-s}\frac{\Gamma(-2k)}{\Gamma(-k)^2}\frac{\Gamma(-s-1)^2\Gamma(s+1-k)}{\Gamma(-s-1-k)}\\
\tilde S_{0k}=(-1)^{k+s+1}\frac{\Gamma(-2k)}{\Gamma(-k)^2}\frac{\Gamma(s+1)^2\Gamma(-k-s-1)}{\Gamma(s+1-k)}
\ee

\subsection*{A2. Explicit calculations for $q\neq 1$}

Similarly to the previous consideration
\be
S_{kk'}=\frac{\int\lm_1^{\infty}d_q x\; x^{k-k'-s-2}\;{}_2\phi_1\left[\begin{array}{c}
		-k\;-k\\
		-2k\\
	\end{array}\right](x^{-1}){}_2\phi_1\left[\begin{array}{c}
	-k'\;-k'\\
	-2k'\\
	\end{array}\right](x)}{\int\lm_1^{\infty}d_q x\; x^{-2(k'+1)}\;{}_2\phi_1\left[\begin{array}{c}
	-k'\;-k'\\
	-2k'\\
\end{array}\right](x)^2}
\ee
where
\be
{}_2\phi_1\left[\begin{array}{c}
	\alpha\; \beta\\
	\gamma\\
\end{array}\right](z)=\sum\lm_{n=0}^{\infty} (-z)^n\frac{\left(\prod\lm_{j=0}^{n-1}[\alpha+j]_q\right)\left(\prod\lm_{j=0}^{n-1}[\beta+j]_q\right)}{\left(\prod\lm_{j=0}^{n-1}[\gamma+j]_q\right)[n]_q!},\\
\left[ n \right]_q=\frac{q^n-q^{-n}}{q-q^{-1}},\\
\int\lm_1^{\infty} d_q x\; x^{-n}=\frac{(-1)^n}{[n-1]_q}
\ee
Formulas of the $q\to 1$ case are generalized straightforwardly:
\be
\int\lm_1^{\infty}d_q x\; x^{-2(k'+1)}\;{}_2\phi_1\left[\begin{array}{c}
	-k'\;-k'\\
	-2k'\\
\end{array}\right](x)^2=\frac{(\left[k\right]_q!)^4 }{\left[2k\right]_q!\left[2k+1\right]_q!}
\ee
and
\be
\tilde S_{k0}=(-1)^{k-s+1}\frac{\Gamma_q(-2k)}{\Gamma_q(-k)^2}\frac{\Gamma_q(-s-1)^2\Gamma_q(s+1-k)}{\Gamma_q(-s-1-k)}\\
\tilde S_{0k}=(-1)^{k+s+1}\frac{\Gamma_q(-2k)}{\Gamma_q(-k)^2}\frac{\Gamma_q(s+1)^2\Gamma_q(-k-s-1)}{\Gamma_q(s+1-k)}
\ee
where,
\be
\Gamma_q(x+1)=\frac{q^x-q^{-x}}{q-q^{-1}}\Gamma_q(x)
\ee

\section*{Acknowledgements}
D.G. would like to thank S.Arthamonov, P.Longhi, G.W.Moore and Sh.Shakirov for valuable and stimulating discussions.
Our work is partly supported by grant
NSh-1500.2014.2, by RFBR  grants 13-02-00457 (D.G. and A.Mir.), 13-02-00478 (A.Mor.),
by joint grants 13-02-91371-ST, 14-01-92691-Ind,
by the Brazil National Counsel of Scientific and
Technological Development (A.Mor.), by the program  of UFRN-MCTI, Brazil (A.Mir.). The work of D.G. is supported by the DOE under grants SC0010008, ARRA-SC0003883, DE-SC0007897.


\begin{thebibliography}{12}

\bibitem{KS} M.Kontsevich and Y.Soibelman, arXiv:0811.2435;\\
See a review and further references in:\\
B. Pioline, 1103.0261

\bibitem{Wit} E.Witten, Comm.Math.Phys. {\bf 121} (1989) 351

\bibitem{RT} E.Guadagnini, M.Martellini and M.Mintchev, In Clausthal 1989,
Proceedings, Quantum groups, 307-317;
Phys.Lett. B235 (1990) 275;\\
N.Yu.Reshetikhin and V.G.Turaev, 
Comm. Math. Phys. {\bf 127} (1990) 1-26

\bibitem{MV} A.Morozov and L.Vinet,
Int.J.Mod.Phys. A13 (1998) 1651-1708, hep-th/9409093

\bibitem{MV2}
A.Mironov, hep-th/9409190; Theor.Math.Phys. {\bf 114} (1998) 127, q-alg/9711006

\bibitem{MMMII} A.Mironov, A.Morozov and And.Morozov, JHEP {\bf 03} (2012)
034, arXiv:1112.2654

\bibitem{UFN3} A.Morozov,
Phys.Usp.(UFN) {\bf 35} (1992) 671-714; {\bf 37} (1994) 1, hep-th/9303139;
hep-th/9502091; hep-th/0502010\\
A.Mironov, Int.J.Mod.Phys. {\bf A9} (1994) 4355, hep-th/9312212; Phys.Part.Nucl.
{\bf 33} (2002) 537

\bibitem{FG} V.V.Fock and A.B.Goncharov, Publ.Math.Inst.Hautes \'Etudes Sci. {\bf 103} (2006) 1-211, math/0311149

\bibitem{CA} S.Fomin and A.Zelevinsky, A.Amer.Math.Soc. {\bf 15} (2002) 497-529, math/0104151;
Composito Math. {\bf 143} (2007) 112-164, math/0602259;\\
S.Fomin, M.Shapiro and D.Thurston, Acta Math. {\bf 201} (2008) 83-146, math/0608367;\\
S.Fomin and D.Thurston, arXiv:1210.5569 [math.GT]

\bibitem{Pop} A.Popolitov,  arXiv:1403.1834

\bibitem{SW} N.Seiberg and E.Witten,
Nucl.Phys., {\bf B426} (1994) 19-52, hep-th/9408099;
Nucl.Phys., {\bf B431} (1994) 484-550, hep-th/9407087

\bibitem{WitMth} E.Witten, Nucl.Phys. {\bf B500} (1997) 3-42,  arXiv:hep-th/9703166

\bibitem{Mikh} A.Mikhailov, Nucl.Phys. {\bf B533} (1998) 243-274, hep-th/9708068

\bibitem{SWint} A.Gorsky, I.Krichever, A.Marshakov, A.Mironov and A.Morozov,
Phys.Lett. {\bf B355} (1995) 466-477, hep-th/9505035;\\
R.Donagi and E.Witten, Nucl.Phys., {\bf B460} (1996) 299-334, hep-th/9510101;\\
H.Itoyama and A.Morozov,
Nucl.Phys., {\bf B477} (1996) 855-877, hep-th/9511125;
Nucl.Phys., {\bf B491} (1997) 529-573, hep-th/9512161;\\
See a review in:\\
A.Gorsky and A.Mironov, hep-th/0011197

\bibitem{Rub} G.F.Bonini, A.G.Cohen, C.Rebbi and V.A.Rubakov, Phys.Rev. {\bf D60} (1999) 076004, arXiv:hep-ph/9901226

\bibitem{AMM1} A.Alexandrov, A.Mironov and A.Morozov,
Int.J.Mod.Phys. {\bf A19} (2004) 4127, hep-th/0310113;
Teor.Mat.Fiz. {\bf 150} (2007) 179-192, hep-th/0605171

\bibitem{AMM2} A.Alexandrov, A.Mironov and A.Morozov, Int.J.Mod.Phys. {\bf A21} (2006) 2481-2518,
hep-th/0412099

\bibitem{AMM3} A.Alexandrov, A.Mironov and A.Morozov, Fortsch.Phys. {\bf 53} (2005) 512-521, hep-th/0412205

\bibitem{GMN2} D.Gaiotto, G.W.Moore and A.Neitzke,
arXiv:0907.3987

\bibitem{GLM} D.Galakhov, P.Longhi and G.W.Moore, 
arXiv:1408.0207

\bibitem{NS} N.Nekrasov and S.Shatashvili,  arXiv:0908.4052

\bibitem{MMS} A.Mironov, A.Morozov, Sh.Shakirov,
JHEP {\bf 02} (2010) 030, arXiv:0911.5721

\bibitem{BS} A.Mironov and A.Morozov, 
JHEP {\bf 04} (2010) 040, arXiv:0910.5670;
J.Phys. {\bf A43} (2010) 195401, arXiv:0911.2396

\bibitem{surop} A.Marshakov, A.Mironov and A.Morozov,  J.Geom.Phys. {\bf 61} (2011)
1203-1222, arXiv:1011.4491

\bibitem{NRS} N.Nekrasov, A.Rosly and S.Shatashvili,  Nucl.Phys. (Suppl.) {\bf B216} (2011) 69-93, arXiv:1103.3919

\bibitem{Krefl} D.Krefl, arXiv:1311.0584; arXiv:1410.7116

\bibitem{AGT} L.Alday, D.Gaiotto and Y.Tachikawa,
Lett.Math.Phys. {\bf 91} (2010) 167-197, arXiv:0906.3219

\bibitem{AGT1} N.Wyllard,
JHEP {\bf 0911} (2009) 002, arXiv:0907.2189;\\
A.Mironov and A.Morozov, Phys.Lett. {\bf B680} (2009) 188-194,
arXiv:0908.2190; Nucl.Phys. {\bf B825} (2009) 1-37, arXiv:0908.2569

\bibitem{GMMMO} A. Gerasimov, A. Marshakov, A. Mironov, A. Morozov, and A. Orlov,
Nucl. Phys.
\textbf{B357} (1991) 565-618;\\
S.Kharchev, A.Marshakov, A.Mironov, A.Orlov and A.Zabrodin,
Nucl.Phys., {\bf B366} (1991) 569-601

\bibitem{MMM} A.Marshakov, A.Mironov, and A.Morozov,
Phys.Lett. {\bf B265} (1991) 99\\
S.Kharchev, A.Marshakov, A.Mironov, A.Morozov and S.Pakuliak,
Nucl.Phys. {\bf B404} (1993) 17-750,  arXiv:hep-th/9208044

\bibitem{GKLMM} A.Gerasimov, S.Khoroshkin, D.Lebedev, A.Mironov, and A.Morozov,
Int.J.Mod.Phys. A10 (1995) 2589-2614, hep-th/9405011

\bibitem{AMM45} A.Alexandrov, A.Mironov and A.Morozov, Physica {\bf D235} (2007) 126-167, hep-th/0608228; JHEP {\bf 12} (2009) 053,
arXiv:0906.3305

\bibitem{NPS} N.Nekrasov, V.Pestun and S.Shatashvili, 
arXiv:1312.6689

\bibitem{CS} S.-S.Chern and J.Simons,
Ann.Math. {\bf 99} (1974) 48-69

\bibitem{WZWN} J.Wess and B.Zumino, Phys.Lett. {\bf B37} (1971) 95;\\
S.Novikov, UMN, {\bf 37} (1982) 37;\\
E.Witten, Comm.Math.Phys. {\bf 92} (1984) 455;\\
A.Gerasimov, A.Marshakov, A.Morozov, M.Olshanetsky, S. Shatashvili,
Int.J.Mod.Phys. {\bf A5} (1990) 2495-2589

\bibitem{KZ} V.Knizhnik and A.Zamolodchikov, Nucl.Phys. {\bf B247} (1984) 83-103

\bibitem{Morton} H.Morton and S.Lukac,
J. Knot Theory and Its Ramifications, {\bf 12} (2003) 395,
math.GT/0108011

\bibitem{GMM} D.Galakhov, A.Mironov and A.Morozov, JHEP {\bf 06} (2014) 050, arXiv:1311.7069

\bibitem{inds} R.K.Kaul and T.R.Govindarajan, Nucl.Phys. {\bf B380} (1992)
293-336, hep-th/9111063;\\
P.Ramadevi, T.R.Govindarajan and R.K.Kaul, Nucl.Phys. {\bf B402} (1993)
548-566, hep-th/9212110;
Nucl.Phys. {\bf B422} (1994) 291-306, hep-th/9312215;\\
P.Ramadevi and T.Sarkar,
Nucl.Phys. B600 (2001) 487-511,
hep-th/0009188;\\
Zodinmawia and P.Ramadevi, arXiv:1107.3918;  arXiv:1209.1346

\bibitem{MMMI} A.Mironov, A.Morozov and An.Morozov, {\sl Strings, Gauge Fields, and the Geometry Behind:
The Legacy of Maximilian Kreuzer,} World Scietific Publishins Co.Pte.Ltd. 2013, pp.101-118, arXiv:1112.5754;\\
H.Itoyama, A.Mironov, A.Morozov, And.Morozov,
Int.J.Mod.Phys. {\bf A27} (2012) 1250099,
arXiv:1204.4785;\\
A.Anokhina, A.Mironov, A.Morozov and And.Morozov, Nucl.Phys. {\bf B868} (2013) 271-313,
arXiv:1207.0279;\\
H.Itoyama, A.Mironov, A.Morozov, And.Morozov,
Int.J.Mod.Phys. {\bf A28} (2013) 1340009,
arXiv:1209.6304;\\
A.Anokhina, A.Mironov, A.Morozov and And.Morozov, arXiv:1304.1486;\\
A.Anokhina and An.Morozov, arXiv:1307.2216

\bibitem{BPZ} A.Belavin, A.Polyakov, A.Zamolodchikov, Nucl.Phys. {\bf B241} (1984) 333-380

\bibitem{PT} B.Ponsot and J.Teschner, 
arXiv:hep-th/9911110;\\
B.Ponsot and J.Teschner, 
Commun.Math.Phys. 224 (2001) 613-655, arXiv:math/0007097

\bibitem{Go} N.Drukker, J.Gomis, T.Okuda and J.Teschner, JHEP {\bf 1002} (2010) 057, arXiv:0909.1105

\bibitem{evo} A.Mironov, A.Morozov and An.Morozov, AIP Conf. Proc. {\bf 1562} (2013) 123, arXiv:1306.3197

\bibitem{RS} M. Rosso and V. F. R. Jones, J. Knot Theory Ramifications, {\bf 2} (1993)
97-112;\\
X.-S.Lin and H.Zheng,
Trans. Amer. Math. Soc. {\bf 362} (2010) 1-18
math/0601267;\\
S.Stevan, Annales Henri Poincare, {\bf 11} (2010) 1201-1224, arXiv:
1003.2861

\bibitem{Kash} R.Kashaev,
Mod.Phys.Lett. {\bf A39} (1997) 269-275

\bibitem{Hikami} K.Hikami, 
Int.J.Mod.Phys. {\bf A16} (2001) 3309-3333, math-ph/0105039;
J.Geom.Phys. {\bf 57} (2007) 1895-1940, math/0604094

\bibitem{Hik1} K.Hikami and R.Inoue, arXiv:1212.6042, arXiv:1304.4776

\bibitem{Hik2} K.Hikami and R.Inoue, arXiv:1404.2009

\bibitem{Guk} T.Dimofte, S.Gukov, J.Lenells and D.Zagier, Commun.Num.Theor.Phys.
{\bf 3} (2009) 363-443, arXiv:0903.2472

\bibitem{CS-cplx} E.Witten, arXiv:1001.2933

\bibitem{Fad} L.Faddeev and R.Kashaev, 
Mod.Phys.Lett. {\bf 9} (1994), 265-282,
hep-th/9310070;\\
L.Faddeev, Lett.Math.Phys. {\bf 34} (1995) 249-254, hep-th/9504111

\bibitem{GMN} D.Gaiotto, G.W.Moore and A.Neitzke, 
arXiv:0807.4723,
arXiv:1006.0146,
arXiv:1103.2598,
arXiv:1204.4824

\bibitem{Ga} L.F.Alday, D.Gaiotto, S.Gukov, Y.Tachikawa and H.Verlinde, 
    JHEP {\bf 1001} (2010) 113, arXiv:0909.0945;\\
D.Gaiotto, 
arXiv:1404.0332

\bibitem{Chekhov} L.Chekhov, B.Eynard and S.Ribault, 
J. Math. Phys. {\bf 54} (2013) 022306, arXiv:1209.3984

\bibitem{Gaiotto} S.Cecotti, D.Gaiotto and C.Vafa, 
arXiv:1312.1008

\bibitem{Cecotti-Vafa} S.Cecotti, A.Neitzke and C.Vafa, 
arXiv:1006.3435

\bibitem{Moore-Seiberg} G.W.Moore and N.Seiberg, Lectures on RCFT, Strings'89, http://www.physics.rutgers.edu/{\textasciitilde}gmoore/LecturesRCFT.pdf

\bibitem{MM} P.Di Francesco, M.Gaudin, C.Itzykson and F.Lesage,
Int.J.Mod.Phys. {\bf A9} (1994) 4257-4352, hep-th/9401163;\\
A.Zabrodin, arXiv:0907.4929;\\
A.Morozov and Sh.Shakirov,
arXiv:1004.2917;\\
A.Mironov, A.Morozov, A.Popolitov and Sh.Shakirov, Theor.Math.Phys. {\bf 171} (2012) 505-522, arXiv:1103.5470;\\
A.Morozov, arXiv:1201.459;\\
A.Mironov, A.Morozov and Z.Zakirova, Phys.Lett. {\bf B711} (2012) 332-335, arXiv:1202.6029

\bibitem{Iorgov} N.Iorgov, O.Lisovyy and Yu.Tykhyy, 
JHEP {\bf 12} (2013) 029, arXiv:1308.4092;\\
N.Iorgov, O.Lisovyy and J.Teschner, 
arXiv:1401.6104

\bibitem{MMZZ}  A.Mironov, A.Morozov, Y.Zenkevich, and A.Zotov,
JETP Letters, {\bf 97} (2013) 45–51, arXiv:1204.0913;\\
A.Mironov, A.Morozov, B.Runov, Y.Zenkevich, and A.Zotov,
Letters in Mathematical Physics, {\bf 103} (2013) 299–329, arXiv:1206.6349;
JHEP {\bf 2013} (2013) 34, arXiv:1307.1502

\bibitem{qT}  R.Kashaev,
Lett.Math.Phys. {\bf 43} (1998) 105–115, q-alg/9705021;\\
V.Fock and L.Chekhov,
Theor.Math.Phys. {\bf 120} (1999) 1245-1259, math/9908165;\\
J.Teschner, math/0510174;\\
J.Andersen and R.Kashaev, arXiv:1305.4291

\bibitem{Aganagic} A.Brini, B.Eynard and M.Mari\~no, arXiv:1105.2012;\\
M.Aganagic, N.Haouzi and S.Shakirov, 
arXiv:1403.3657;\\
A.Alexandrov, A.Mironov, A.Morozov and An.Morozov, JETP Letters, {\bf 100} (2014) 271-278, arXiv:1407.3754

\bibitem{Olshanetsky} M.Olshanetsky, 
    Lett.Math.Phys. {\bf 42} (1997) 59-71, arXiv:hep-th/9510143

\bibitem{mamoCFT} Vl.Dotsenko and V.Fateev, Nucl.Phys. {\bf B240} (1984) 312-348;\\
R.Dijkgraaf and C.Vafa, arXiv:0909.2453;\\
H.Itoyama, K.Maruyoshi and T.Oota,
Prog.Theor.Phys. {\bf 123} (2010) 957-987, arXiv:0911.4244;\\
T.Eguchi and K.Maruyoshi,
arXiv:0911.4797;
arXiv:1006.0828;\\
R.Schiappa and N.Wyllard,
arXiv:0911.5337;\\
A.Mironov, A.Morozov, Sh.Shakirov,
Int.J.Mod.Phys. {\bf A25} (2010) 3173-3207,
arXiv:1001.0563;  JHEP {\bf 1102} (2011) 067,
arXiv:1012.3137;\\
A.Mironov, A.Morozov and And.Morozov,
Nucl.Phys. {\bf B843} (2011) 534-557, arXiv:1003.5752

\bibitem{GMM1} D.Galakhov, A.Mironov and A.Morozov, 
JHEP 2012 (2012) 67, arXiv:1205.4998

\bibitem{Nem} N.Nemkov,  J.Phys. A: Math. Theor. {\bf 47} (2014) 105401,  arXiv:1307.0773; arXiv:1409.3537

\bibitem{Zam} A.Zamolodchikov and Al.Zamolodchikov,
\emph{Conformal field theory and critical phenomena in 2d systems}, 2009 (in Russian)

\bibitem{large_c} A.Mironov and A.Morozov, Phys.Lett. {\bf B682} (2009) 118-124, arXiv:0909.3531

\bibitem{OEIS} The On-Line Encyclopedia of Integer Sequences at http://oeis.org/

\end{thebibliography}
\end{document}